\documentclass[letterpaper,twocolumn,10pt]{article}
\usepackage{usenix-2020-09}

\usepackage{tikz}
\usepackage{amsmath}

\usepackage{filecontents}
\usepackage{cite}
\usepackage{amsmath,amssymb,amsfonts}
\usepackage{graphicx}
\usepackage{textcomp}
\usepackage{xcolor}
\def\BibTeX{{\rm B\kern-.05em{\sc i\kern-.025em b}\kern-.08em
    T\kern-.1667em\lower.7ex\hbox{E}\kern-.125emX}}

\usepackage{amsthm}
\usepackage{url}
\numberwithin{equation}{section}
\usepackage[draft]{todonotes}
\usepackage{xspace}
\usepackage[ruled,vlined,algo2e]{algorithm2e}
\usepackage{algorithm}
\usepackage{booktabs} 
\usepackage{wasysym}
\usepackage{physics}

\newtheorem{theorem}{Theorem}
\newtheorem{lemma}{Lemma}
\newtheorem{definition}{Definition}

\newcommand{\threevdots}{%
  \vbox{\baselineskip1ex\lineskiplimit0pt%
  \hbox{.}\hbox{.}\hbox{.}}}

\newcommand{\SubXBox}{\scalebox{.76}{\XBox}}
\newcommand*{\thead}[1]{\multicolumn{1}{c}{\bfseries #1}}

\definecolor{forestgreen(web)}{rgb}{0.13, 0.55, 0.13}

\newcommand\sysname{\textsc{VAMS}\xspace}

\newcommand{\request}{\mathit{request}}
\newcommand{\provide}{\mathit{provide}}
\newcommand{\ucheck}{\mathit{check}}
\newcommand{\monitor}{\mathit{monitor}}
\newcommand{\udetect}{\mathit{detect}}
\newcommand{\audit}{\mathit{audit}}
\newcommand{\publish}{\mathit{publish}}
\newcommand{\odetect}{\mathit{detect}}
\newcommand{\host}{\mathit{host}}
\newcommand{\broker}{\mathit{broker}}

\newcommand{\ida}{\mathit{id_a}}
\newcommand{\idp}{\mathit{id_{dp}}}
\newcommand{\idc}{\mathit{id_c}}

\newcommand{\sid}{\mathit{n}}

\begin{document}
\date{}

%\subtitle{Subject Section}

\title{\Large \bf VAMS: Transparent Auditing of Access to Data}

%for single author (just remove % characters)
\author{
{\rm Alexander Hicks}\\
University College London
\and
{\rm Vasilios Mavroudis\thanks{Work done while at University College London}}\\
Alan Turing Institute
\and
{\rm Mustafa Al-Bassam\footnotemark[1]}\\ 
Celestia
\and 
{\rm Sarah Meiklejohn} \\
University College London \& Google
\and 
{\rm Steven J. Murdoch} \\
University College London
% copy the following lines to add more authors
% \and
% {\rm Name}\\
%Name Institution
} % end author

% \author[1,$\dagger$]{Alexander Hicks}
% \author[2]{Vasilios Mavroudis\thanks{Work done while at University College London}}
% \author[3]{Mustafa Al-Bassam\footnotemark[1]}
% \author[1,4]{Sarah Meiklejohn}
% \author[1]{Steven J. Murdoch}

% \authormark{Author Name et al.}

% \address[1]{\orgname{University College London}}
% \address[2]{\orgname{Alan Turing Institute}}
% \address[3]{\orgname{Celestia}}
% \address[4]{\orgname{Google}}

\maketitle

\begin{abstract}
We propose \sysname, a system that enables transparency for audits of access to data requests without compromising the privacy of parties in the system.  
\sysname supports audits on an aggregate level and an individual level, by relying on three mechanisms.
A tamper-evident log provides integrity for the log entries that are audited.
A tagging scheme allows users to query log entries that relate to them, without allowing others to do so.
MultiBallot, a novel extension of the ThreeBallot voting scheme, is used to generate a synthetic dataset that can be used to publicly verify published statistics with a low expected privacy loss.
We evaluate two implementations of \sysname, and show that both the log and the ability to verify published statistics are practical for realistic use cases such as access to healthcare records and law enforcement access to communications records.
\end{abstract}

% \keywords{transparency, privacy, data access requests} % TODO: replace with your keywords

%%%%%%%%%
\section{Introduction}\label{sec:introduction}
Personal data plays an important role in activities where there is a high cost of failure e.g., health-care, preventing and detecting crime, and legal proceedings.
Often, however, the organizations that need access to this data are not the ones who generate or hold the data, so data must be shared for it to be used.
Such sharing must be done with care as improper sharing or modification of sensitive data can result in harm to the individuals whose data is involved, and others, whether through breaches of confidentiality or incorrect decisions as a result of tampered data.
If there is widespread abuse of personal data, people may become unwilling to allow their data to be collected and processed even when it would benefit themselves and society.

Simple restrictions on sharing of personal data can be automatically enforced through access control and cryptographic protections, such as preventing unauthorized parties from accessing databases in which personal data is held.
However, other equally important restrictions involve human interpretations of rules, consent, or depend on information not available to the computer system enforcing them.
For example, access to medical records may be permitted only when it would be in the interests of the patient.
Similarly, access to communication records may be permitted only if it is necessary and proportionate to prevent crime.

In such cases, rules cannot reliably be automatically enforced in real-time so the approach commonly taken is to keep records of access attempts and subject the actions to audit.
Provided that the audit can detect improper activities and violations are harshly punished, abuse can be effectively deterred.
Statistics published about the audit can also provide confidence to society that access to data is being controlled and that organizations who can access data will be held to account.

This raises questions about who performs the audit and how the auditor can be assured that the records they see are accurate.
If individuals at risk of their personal data being misused do not trust that the auditor is faithfully carrying out their duties then the goal of the audit will not be achieved.
However, because of the sensitivity of personal data and the records containing the justification for data being processed, not everyone can act as an auditor.
Even if it was possible to find an organization whose audit would be widely accepted, an audit based on tampered records would not be reliable.

The integrity of the data that is accessed is also important when actions are taken based on this data.
When making a medical decision or conducting legal proceedings, relying on tampered data can have severe consequences.
It may be possible to refer back to the organization that collected the data to verify its integrity, but if that organization no longer holds the data or has gone out of business, such verification is not possible.
Digital signatures can provide some confidence that data is genuine, but if the private key is compromised then any data signed by that key is subject to doubt, even if it was created before the point of key compromise.

To improve on the current situation, we propose \sysname, a system that enables transparent audits of access to data requests.
This is achieved by allowing auditors to verify the integrity of the data they see and publish audits that can be publicly verified without compromising the privacy of the parties involved, as well as allowing individuals to audit requests for data that relates to them.

\section{Outline of the paper}
Section~\ref{sec:threat} introduces our setting, threat model, and goals, which address transparency (verifiable audits of aggregate and individual outcomes) and privacy (the verifiability of audits cannot reveal more than what is intentionally revealed by audits).

We describe in Section~\ref{sec:blocks} the three mechanisms that we use to build \sysname.
Tamper-evident logging provides integrity for the information they see on the log.
A log entry tagging scheme allows users to efficiently find log entries that are relevant to them.
MultiBallot, a novel adaptation of ThreeBallot~\cite{rivest2007three} as a rule-based way of generating a synthetic dataset, allows published audits to be publicly verified with only a small expected privacy loss.

The operation of \sysname is described in Section~\ref{sec:system}, while Section~\ref{sec:security} argues that it achieves the goals stated in Section~\ref{sec:threat}, and Section~\ref{sec:implementation} shows that the two implementations of the log, based on Hyperledger Fabric (HLF) and Trillian, show sufficient scalability and functionality, as well as the ability to accurately verify statistics with MultiBallot.
Our results show that \sysname can serve as a lightweight overlay applicable to many use cases.

\section{Motivating scenarios}\label{sec:background}

To motivate the design of our system, we consider two challenging scenarios: controlling the access of law-enforcement personnel to communication records and the access of healthcare professionals to medical data. 

\subsection{Law-enforcement access to communications data}\label{sec:law-enforcement}

In the UK 95\% of serious and organized crime cases make use of communications data~\cite{communicationsdata} -- metadata stored by telecommunications providers in their billing system about account holders or their use of communications networks (e.g., phone numbers called, address associated with an account, location of a mobile phone).
Telecommunications providers are required to store this data for up to 2 years, but once this period has expired and there is no business reason to store this personal data, they are required to delete it.
Within the period that data is stored, law-enforcement personnel is permitted to request access, provided that they can demonstrate that their actions are legally justified\footnote{Similar legal powers are available in the US through the use of administrative subpoenas, but as there are no publicly available statistics for their use and there is no centralized oversight, we focus on the UK case.}.
At the time a request is made, there is, however, no external oversight.
Instead, information about the request and the justification for access are stored and made available for audit by the Investigatory Powers Commissioner's Office (IPCO)\footnote{Before to September 2017 this role of IPCO was the responsibility of the Interception of Communications Commissioner's Office (IOCCO).}.
IPCO then assess whether law enforcement personnel make appropriate use of the powers they were given, and publishes reports with statistics of how these powers were used~\cite{iocco2016}.

Communications data plays an important role in the investigation of criminal offences, but may also be used as evidence in legal proceedings, for the prosecution or defence.
If the integrity of the evidence is questioned, a representative of the telecommunications provider will be asked to appear in court to verify the evidence and attest to its accuracy.
If technical issues arise related to this evidence, one of the parties to the case may also request that the court request assistance from an expert witness.
This process is expensive, time-consuming, and even impossible if the provider has deleted the original data in the time between the law enforcement agency requesting it and the data being required in court.

To improve the process, industry standards allow providers to sign or hash communications data when it is provided in response to a request from law enforcement.
Someone who needs to verify an item of data can compare the hash to the one stored by the provider, or verify the digital signature using the provider's public key~\cite{ETSI}.
However, if the provider's private key or hash database is compromised, any evidence presented after to the compromise will be brought into doubt, even if it was generated before the time of compromise.

Our system can be applied in this scenario, allowing the integrity of communications data evidence to be demonstrated, even if the communications provider which produced the data no longer exists or has been compromised.
Furthermore, the system will give assurance to the auditor that records of requests to access communications data have not been tampered with, and assure society that reported statistics have not been improperly manipulated by the auditor.
We also show how the system protects the privacy of individuals whose data is requested and also protects the confidentiality of ongoing law-enforcement investigations.

% \alex{new ETSI standard? ETSI TS 103 307 V1.4.1 (2021-06)}

\subsection{Access to healthcare records}
In our second scenario, we consider how to empower individuals by giving them control over how their medical records are used and shared.
In a healthcare system, once consent has been given by a patient, various actors should be able to access various records associated with that patient;
e.g., their general practitioner should be able to access scans that were run at a hospital, and researchers running academic studies or clinical trials in which the patient has enrolled should be able to access records relevant to the study.

Currently, patients can only give permission for broad types of activities and may have legitimate concerns that their information is being used inappropriately.
Conversely, patients with serious diseases (e.g., cancer, motor neuron disease) often have trouble getting the treatment they need, as universities conducting studies are legally blocked from contacting them, and patients are unaware that such studies are going on.

Opening up access to medical databases may fulfil the needs of some patients but would also open up the potential for abuse, so it is important for patients to have visibility into how their data is being used to understand the implications of their consent.
For clinical practice, the default could be that patients opt-in to sharing their data, although they can always opt out if they wish.
For academic studies and clinical trials, the default should be that they are opted out, but can opt-in.
They can even choose at some granular level (e.g., according to the type of study) which studies they want to opt in to.

One issue with having patients opt in individually is that for some studies this process may not result in a large enough sample.
Equally, if patients are deluged with requests for consent, they are likely to resort to some default behaviour (``click-through syndrome'') without understanding what they have consented to.
As such, patients could outsource these decisions to data brokers; i.e., organizations that pay attention to the studies being conducted and are authorized to provide consent on behalf of patients registered with them.

Our system can be applied to allow patients to share their data in such a way as to protect their privacy while ensuring that unauthorized parties are prevented from having access and that authorized parties abusing their access can be detected.

%%%%%%%%%
\section{Threat model and Goals}\label{sec:threat}
% \subsection{Setting}
\begin{figure}[t]
\centerline{\includegraphics[width=\linewidth]{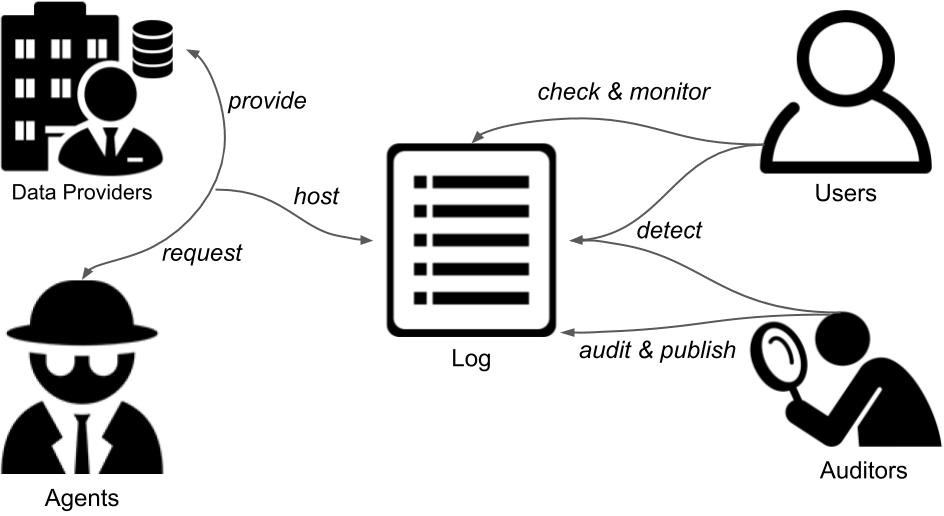}}
  \caption{The parties in \sysname and their functionalities. The optional data broker would act as a user.}
  \label{fig:setting}
\end{figure}

Our setting (illustrated in Figure~\ref{fig:setting}) involves agents, data providers, users, auditors, log servers, and (optional) data brokers.
Table~\ref{table:functions} summarises the functionalities and malicious behaviours for each party, which we describe below.
%External to the system are also regulators that determine the policies under which the parties in the system operate.

\begin{table*}[t]
\begin{center}
\caption{The parties in the system, the functions they perform and their malicious behaviour.}
\label{table:functions}
\resizebox{\linewidth}{!} {%
\begin{tabular}{l l l}
\thead{Party} & \thead{Function} & \thead{Malicious behaviour} \\
\midrule
Agent & $request$: append a request to a data provider to the log & Provide an invalid request \\
Data provider & $provide$: answer a request that is on the log & Provide invalid data \\
& $detect$: detect if log servers are behaving dishonestly & \\
User & $check$: look for relevant log entries & Access requests relevant to other users \\
& $monitor$: verify the statistics published by auditors & Infer information from the statistics \\
& $detect$: detect if log servers are behaving dishonestly & \\
Auditor & $audit$: check the log entries for misuse or errors & Infer information from the log \\
& $publish$: publish statistics about entries on the log & Publish inaccurate statistics \\
& $detect$: detect if log servers are behaving dishonestly & \\
Log server & $host$: return the log to parties wishing to inspect it & Provide inconsistent views of the log \\
Data broker & $broker$: respond to requests in place of a user according to their preferences & Misrepresent the preferences of the user \\
\end{tabular}
}
\end{center}
\end{table*}

The \textit{log} is a key-value store of access to data requests.
The values of log entries are \textit{records} i.e., tuples of \textit{elements} such as the attributes of a data request (e.g., the type of data requested by law enforcement) or answers to a medical questionnaire.
The log can also contain datasets and statistics published by auditors or a link to the datasets and statistics along with a hash to verify their integrity.
\textit{Log servers} host the log of requests ($\host$).
A malicious log server would aim to give inconsistent views of the log to auditors and users i.e., attack the availability of logged information.

\textit{Agents} (e.g., law enforcement, medical researchers) request access to user data from data providers ($\request$).
A malicious agent would aim to access data without it being logged or submit an invalid request, and try to tamper with a logged invalid request before an auditor or user audits it.
In other words, a malicious agent would aim to attack the integrity of the log.

\textit{Data providers} (e.g., telecommunications providers, healthcare providers, users) collect user data and receive requests from agents ($\provide$).
A malicious data provider would aim to give access to data without it being logged.

\textit{Auditors} audit the log ($\audit$) and publish statistical reports ($\publish$).
In the UK, the IPCO is an example of this kind of auditor.
They must be able to detect if log servers are behaving dishonestly ($\odetect$).
A malicious auditor would aim to publish an inaccurate report i.e., compromise the integrity of the audit.

\textit{Users} are members of the public that check requests for their data ($\ucheck$) and verify audits ($\monitor$).
They must also be able to detect if log servers are behaving dishonestly ($\odetect$).
A malicious user would aim to learn information about another user (i.e., attack their privacy) by using the log or published audits.
% through the log e.g., access requests made about other users or infer information from published reports.

\textit{Data brokers} are non-essential intermediaries that users can rely on to deal with requests if they are willing to serve as a data provider
e.g., by providing data to a study.
A data broker can then deal with requests ($\broker$) according to pre-set rules from the user.
A malicious broker would aim to misrepresent the user's preference e.g., accept requests that the user's rules would prohibit.

\subsection{Threat Model}
We allow every party to act maliciously except for colluding data providers and agents as they could simply exchange data without it being logged.
% This is because \sysname requires any access to data request to be logged.
% If a data provider and an agent collude they could simply exchange the data without it being logged.
This cannot be prevented with cryptographic techniques because the data must generally exist in an unencrypted form for its primary use e.g., routing calls or providing healthcare.
% by \sysname or any other similar system.
We, therefore, require that agents log their requests and that data providers do not answer requests without ensuring that the request has been logged.
If one of these parties is malicious, their misbehaviour will be caught by the other.

% Cryptographic mechanisms like those discussed in this paper cannot protect against this scenario because the data must generally exist in an unencrypted form for its primary use (e.g., routing calls, providing healthcare).
In practice, \sysname as described here would be augmented by procedural and technical access controls that prevent confidential data from leaving a system without being logged.
The goal of \sysname is to help ensure that requests represented by the log are compliant with policy and have not been tampered with.

% \subsection{Transparency and privacy goals}\label{sec:goals}
% % \sysname aims to offer transparency about requests for access to data and related audits, and ensure the privacy of parties in the system.
% Our motivation is to provide more agency to users, who in many current systems have data requested and provided about them but cannot evaluate this process.
% Transparency can help resolve this asymmetry, but it must be balanced with privacy goals that ensure parties do not gain more sensitive information about others.
% % than they would have had otherwise.
% Table~\ref{table:goals} summarises our transparency and privacy goals, which are explained below.

\begin{table*}[t]
\begin{center}
\caption{Transparency and privacy goals that address the malicious behaviours defined in our threat model.}
\label{table:goals}
\resizebox{\linewidth}{!} {%
\begin{tabular}{l l l}
\thead{Goal} & \thead{Supports} &\thead{Protects against} \\
\midrule
Log availability (T1) & Agents and users ($\udetect$)& Log servers providing inconsistent views of the log \\
Log integrity (T2) & Agents ($\audit$) and users ($\ucheck$) & Agents tampering requests \\
Verifiability of inputs used to compute statistics (T3) & Users ($\monitor$) & Auditors releasing inaccurate statistics \\
Verifiability of published statistics (T4) & Users ($\monitor$) & Auditors inaccurate statistics \\
Transparency of the system (T5) & Auditors and users & Reliance on agents, data providers, or (for users) auditors \\
The log itself does not reveal any sensitive information (P1) & User, agent, data provider privacy & Parties wanting to infer information from the log \\
Verifying an audit is privacy preserving (P2) & User privacy & Parties wanting to infer information from the statistics \\
\end{tabular}
}
\end{center}
\end{table*}

\subsubsection{Transparency goals}\label{sec:transparencygoals}
% \alex{emphasize editorial control/indidividual evidence?}
Our motivation is to provide more agency to users, who in many current systems have data requested and provided about them but cannot evaluate this process.
To help resolve this asymmetry, \sysname is designed to provide verifiable aggregate statistics (population outcomes) about requests for data and the use of data that can be verified by users, and allow users to check for log entries that are relevant to them (individual outcomes).
This is achieved through five transparency goals, which are summarised in Table~\ref{table:goals} .

\paragraph{Log availability (T1).} It must be possible for users and auditors to access the information they require for their respective audits.

\paragraph{Log integrity (T2).} Users and auditors must also be able to check the integrity of the information they access because the information on the log could have been modified or they could be given an incorrect view of the log.

\paragraph{Verifiability of inputs used to compute statistics (T3) and of published statistics (T4).} Due to the sensitivity of personal data and the records containing the justification for data being processed, not everyone can act as an auditor with wide-ranging access to log entries.
Auditors are therefore relied on to compute and publish aggregate statistics.
To minimise the trust required in the auditors, users must be able to verify both the input and the output of the computation of the statistics. 
Auditors could otherwise publish bogus statistics by miscomputing them or by computing them on a fake dataset that gives the results they desire.

\paragraph{Transparency of the system (T5).} Users and auditors should only have to rely on \sysname itself to perform their functions and not a potentially malicious party.

\subsubsection{Privacy goals}\label{sec:privacygoals}
Our transparency goals must be complemented by \emph{privacy} goals to ensure that parties in the system do not learn private information about one another in the process of performing their audits.
This may seem contradictory, but information relating only to a single party is not necessarily required to evaluate the system as a whole.
\sysname does not control the contents of a published audit so it cannot control the privacy loss associated with an audit, which will vary based on the requirements and privacy concerns of auditors.
As a result of this, our privacy goals (summarised in Table~\ref{table:goals}) focus on requiring that \sysname does not lead to a greater loss of privacy than what is released through a published audit.

\paragraph{The log itself does not reveal any sensitive information (P1).} This requires that the log entries themselves do not reveal any sensitive information, but also that the log as a whole does not reveal links between requests so the log entries should be unlinkable.

\paragraph{Verifying an audit is privacy-preserving (P2).} It should not be possible to learn information about individuals from statistics published by an auditor i.e., verifying an audit should not reveal more than the correctness of the audits themselves, although the audit itself may reveal sensitive information. 

% Our first privacy goal, $P1$, is that the log itself should not lead to any privacy loss.
% This requires that the log entries themselves do not reveal any sensitive information, but also that the log as a whole does not reveal links between requests so the log entries should be unlinkable.
% Our third privacy goal, $P3$, is that verifying the statistics (their inputs and the computed statistics) published with an audit is privacy-preserving i.e., it should not reveal more than the correctness of the audits themselves. 
% This ensures that nothing more than the results of an audit is learned, although the audit itself may reveal sensitive information.

% Our privacy goals are that (1) it should not be possible for a user to gain information about another user by running $\ucheck$ as them, (2) it should not be possible to learn information about individuals from statistics published by an auditor, and (3) the log must also protect the sensitive information that could be part of a logged request.

%%%%%%%%%
\section{Building \sysname}\label{sec:blocks}
\sysname is built using technical mechanisms that we describe below, based on the rationale that follows.

% Before describing and evaluating our proposed system, we introduce the building blocks that will be used to construct it.
% Specifically, we introduce the primitives and mechanisms that will ensure we meet the privacy and transparency goals we have laid out in Section~\ref{sec:threat}.

Central to \sysname is the log containing requests submitted by agents through $\request$.  
For the log, key-value stores are a natural choice as log entries (key-value pairs) include keys (i.e., identifiers) that can easily be queried by users performing $\ucheck$.
Log integrity requirements mean that auditors and users should be able to verify that the records that they access are those submitted by agents, so the log must be tamper-evident.

Auditors and users should also be able to detect log misbehaviour e.g., equivocation in the form of an altered log or a split-view attack in which different versions of the log to different parties.
The need for a tamper-evident log can also be seen in cases where evidence is required e.g., when an agent must show that they accessed data with a valid request.
In some cases, \textit{urgent} requests that are authorized orally (with paperwork authorized only retroactively) are necessary e.g., in the case of a medical or security emergency, so attempting to block invalid requests as they are issued is not enough.

Requests should be signed so that they can be used as evidence to assign liability and to hold the relevant parties accountable.
This would work only if the evidence produced is robust so that liability can be properly assigned.
Evidence should also exist even if the party that produced it is no longer active e.g., if a data provider declares bankruptcy, or if some servers fail, or are destroyed. 
Thus, logs should not depend solely on the party tied to the evidence.

Once requests are recorded in the log, auditors perform their audits and publish the resulting statistics through $\publish$.
Users must be able to verify these statistics through $\monitor$ (requiring the published statistics and the data necessary to verify their results) without learning more than what is revealed by the statistics themselves i.e., specific information about other individual users.
% Users must also be able to check that their data was included in the computation of the statistics to ensure that auditors have not simply generated data that fits their statistics.

To summarize, we need the following.
First, we need some kind of tamper-evident log, which requires that the state updates of the log be tied to a blockchain, a history tree~\cite{crosby2009efficient}, or more generically a transparency overlay~\cite{chase2016transparency} with efficient proofs of inclusion and consistency.
Second, we need a mechanism that allows the log to be efficiently queried (i.e., identifying relevant requests) without revealing any links between entries on the log.
Third, we need a mechanism to publish statistics that can be verified without revealing more information than what can be learned from the statistics themselves.

\subsection{Using Hyperledger Fabric and Trillian as tamper-evident logs}
Existing transparency overlays come in the form of distributed ledgers and verifiable logs.
We have implemented \sysname twice, using Hyperledger Fabric, a distributed ledger with an underlying blockchain, and Trillian, a verifiable log-backed map.

In both cases, using HLF or Trillian guarantees that the log is tamper-evident due to the underlying blockchain or verifiable log that records state updates, that the availability of information on the log can be assured by making log equivocation detectable, and that the log is easy to query as it is in the form of a key-value store.

\paragraph*{Hyperledger Fabric}
Hyperledger Fabric~\cite{hyperledger,hlf-architecture,vukolic2017} is a modular open-source system for deploying and operating permissioned distributed ledgers whose state updates are recorded on a blockchain.

A HLF network is composed of peers, who maintain a key-value store that is updated through transactions on the underlying blockchain, and an ordering service (i.e., a consensus protocol).
Because updates to the ledger's state (i.e., \sysname's log) are recorded on the underlying blockchain that is append-only, the ledger's state benefits from availability guarantees against the log equivocating as a log server that equivocates results in a fork of the blockchain.
Integrity guarantees against tampering of the log are also guaranteed as changes to the ledger's state appears on the blockchain, which can be replayed or queried through a key history function.

% Peers have identities in the form of X.509 public-key certificates~\cite{yee2013updates} and a \textit{Membership Service Provider} (a PKI).
Peers have identities in the form of X.509 public-key certificates and a \textit{Membership Service Provider} (a PKI).
These identities allow peers to be split up into organizations on the network (e.g., agents, data providers, \ldots).
This provides a way of implementing basic access control for operations on the network.

Updating the state of the ledger requires \textit{endorsing peers} to execute \textit{chaincode} (smart contracts) and sign the transaction containing the resulting state update.
This ensures that an endorsing peer can be held accountable for the transactions they endorse e.g., the requests they make as agents or the requests they accept as data providers.

Transactions are then sent to the ordering service that packages them into blocks with which \textit{validating peers} update the state of the ledger.
Only endorsing peers are required to execute code for a transaction, so other peers do not handle any computational burden other than receiving events from the network.
The endorsement mechanism also allows for endorsement policies that limit which peers can invoke or sign transactions for a certain chaincode, for example, based on their organization.

% Because updates to the ledger's state (i.e., \sysname's log) are recorded on the underlying blockchain that is append-only, the ledger's state benefits from availability guarantees against the log equivocating, as a log server that equivocates results in a fork of the blockchain, and integrity guarantees against tampering of the log, as changes the ledger's state appears on the blockchain, which can be replayed or queried through a key history function.

\paragraph*{Trillian}
Trillian~\cite{trillian} is an open-source project that implements a generalization of Certificate Transparency~\cite{CT} based on three components: a verifiable log, a verifiable map, and a log of map heads.

Trillian's verifiable log (not to be confused with \sysname's log) is an append-only log implemented as a Merkle tree that allows clients to efficiently verify that an entry is included in the log (with a proof showing the Merkle path to the tree's entry), detect log equivocation (i.e., conflicting tree heads), and verify that the log is append-only (through Merkle consistency proofs).

The verifiable map (i.e., \sysname's log) is a key-value store implemented as a sparse Merkle tree pre-populated with all possible keys as leaves e.g., all $2^{256}$ possible SHA-256 hashes.
Although a tree with $2^{256}$ unique leaves would not be practical to compute, only the non-empty leaves have to be computed because all others will have the same value (e.g., zero)~\cite{laurie2012revocation}.
Clients can then verify that a certain value is included (or not) in the map at any point in time, with proofs containing Merkle paths.
Combining a verifiable log with a verifiable map leads to a verifiable log-backed map, where the log contains an ordered set of operations applied to the map.
Clients can then verify that the entries in the map they view are the same as those viewed by others by replaying the log and detecting any change in values of the key-value store.

The log of map heads records the root hash of the log, signed by the log's server so that if it equivocates there is a cryptographic proof that it has done so.
Because Trillian does not involve a consensus protocol, it instead relies on gossip between clients (e.g., auditors and users performing $\udetect$) to detect misbehaving servers by comparing the views of the log that they have received.

As in the case of HLF, the fact that updates to the verifiable map (\sysname's log) are recorded on Trillian's append-only verifiable log provides availability guarantees again \sysname's log equivocating as this will lead to different three heads in the log of map heads, and integrity guarantees against tampering of \sysname's log as updates will appear on the underlying append-only Merkle tree that is Trillian's log.

\subsection{Tagging log entries with common identifiers}\label{subsec:tags}
In order to gain useful information from \sysname, users must be able to efficiently identify the log entries that contain information that is relevant to them.
The values of log entries (i.e., records) will be encrypted for privacy reasons so this requires a mechanism that allows agents and data providers to derive keys that users can also compute only if the entry is relevant to them.

A strawman solution would require users to ask agents for the log entries relevant to them.
This would reveal to agents which users are monitoring the log, and require users to trust agents that could lie about the log entries that are relevant to them.
To remove the need for users to interact with agents, we use \textit{common identifiers} that can be computed only with information known to a user, the data provider, and the agent that is involved in a request.
By relying on shared information we remove the need for interaction, and by having the information known to not only the agent and user but also the data provider, a data provider that is not colluding with the agent can check that the tag will be correct.

% We assume that agents and data providers refer to a given user using (private) agent and data provider identifiers, $\ida$ and $\idp$, which are known to the user. 
% It is then possible to obtain common identifiers $\idc=\ensuremath{Enc}(\ida,\idp\|\sid)$ e.g., by using AES, where $\sid$ is a \textit{session identifier} that changes deterministically with every request involving the same pair $(\ida,\idp)$.
% This ensures different requests involving the same parties are unlinkable, but allows users to check relevant requests without communicating with agents or data providers, which also reduces the risk of information leaking.
% The private contents of the requests are then simply encrypted under the keys of users and auditors.

We assume that agents and data providers refer to a given user using (private) agent and data provider identifiers, $\ida$ and $\idp$, which are also known to the user they correspond to but not to others. 
It is then possible to obtain common identifiers $\idc=Hash(\ida\|\idp\|\sid)$ by using a secure hash function such as SHA-256, where $\sid$ is a \textit{session identifier} that changes deterministically with every request involving the same pair $(\ida,\idp)$.

% \alex{Long term concern for security of hash function cf what bitcoin does with double hashing. Could do SHA2(SHA3(inputs)) or something similar.}

This ensures different requests involving the same parties are unlinkable as common identifiers will appear random, but allows users to check requests that are relevant to them communicating with agents or data providers, which also reduces the risk of information leaking.
The private contents of the requests are then simply encrypted under the keys of users and auditors so that they can access them.
% In cases like law enforcement, it may not be appropriate for a user to be able to check for requests as they are submitted (e.g., during an ongoing investigation) but a commitment can be logged instead (without a common identifier) to timestamp 

We assume that $\ida$ and $\idp$ are pseudorandom strings like, for example, the German Electronic Identity Card that can be used for online authentication and has been analysed from a cryptographic point of view~\cite{dagdelen2013cryptographic}.
This is not unreasonable for purpose-built identifiers, and although it adds the burden of managing them, software such as password managers or data brokers could be relied on.
% If $\ida$ and $\idp$ have low entropy, a key derivation function such as the standardized PBKDF2~\cite{pbkdf2} or the more modern Argon2~\cite{biryukov2016argon2} could be used. 
% \alex{THIS IS NOW IN PLACE - Some review mentioned using SHA instead of AES. This seems to make sense - using AES would allow someone that knows only the identifier used as key to decrypt the tag and recover the other identifier. The only downside would be if a hash would be more susceptible to bruteforce but it doesn't seem like that would be the case.}

An agent or data provider could in principle leak $\ida$, $\idp$, or common identifiers, but they could just as well leak the data attached to it in the first place.
% as they control it, and they do not a priori have anything to gain from doing so.
As we have remarked in the threat model, this cannot strictly be prevented as they control the data much like they must know the common identifiers to tag the log entries.
Moreover, they would not gain anything from doing so.

\subsection{Generating synthetic data and verifying statistics with MultiBallot}\label{subsec:multiballot}
To allow published statistics to be publicly verified without incurring more of a privacy loss than is already caused by the statistics themselves, we introduce a way for auditors to generate a synthetic dataset $D_{priv}$ from the dataset $D$ used to compute the statistics.
We call this randomized mechanism MultiBallot and denote $\mathcal{M}_n:D\mapsto D_{priv}$ the mapping of $\vert D \vert$ records in D to $n\vert D \vert$ shares in $D_{priv}$.

\begin{figure}[t]
\centerline{\includegraphics[width=\linewidth]{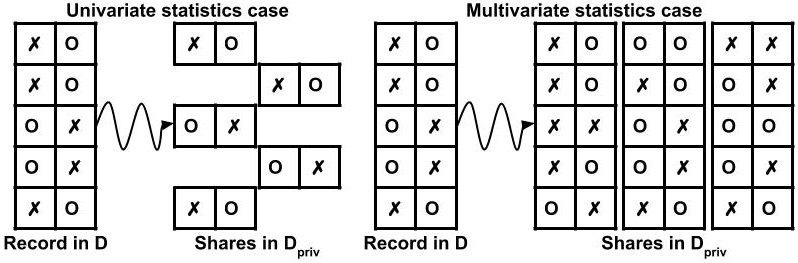}}
  \caption{Example transformation of records in $D$ to shares in $D_{priv}$ for univariate and multivariate statistics. In the univariate case, the record is split into individual elements. In the multivariate case, the record is used to generate shares with the same number of elements that are then split from each other.}
  \label{fig:multiballot}
  \label{fig:threeballot}
\end{figure}

MultiBallot can be used to support either exclusively univariate statistics or multivariate statistics.
Figure~\ref{fig:multiballot} illustrates how a record in a dataset $D$ that has $e$ binary elements can be transformed into \textit{shares} of a synthetic dataset $D_{priv}$ according to the rules we describe for each case.
Elements, in this case, could be attributes of a request for data in the law enforcement case e.g., ``request type:urgent/not urgent'', or of a patient in a medical study e.g., ``has gene X:yes/no''.

% inspired by Rivest's ThreeBallot voting scheme~\cite{threeballot,rivest2007three}, which works by giving voters three ballots instead of one, and having them fill them according to set rules.

% Each row (i.e., element) must be selected at least once and no row must be selected thrice, selecting a row in two columns is a vote for and selecting a row in one column is a vote against.
% The outcome is the same as that of a standard election, and voters can use receipts for one of their ballots to check that their vote was counted (assuming a public bulletin board) and ballots are unlinkable, preventing ballot triplets from being reconstructed.

% We retain the idea of having multiple ballots for each ``vote'', which will be replaced by attributes of data in our context, and use this to generate a differentially private synthetic dataset.
% We generalise the scheme to any odd number $n$ of ballots rather than restricting it to three, which allows for more control over the level of privacy as it allows for a greater number of possible ballots.

% In order to fill in a 

% Figure~\ref{fig:multiballot} illustrates how a record in a dataset $D$ that has $e$ binary elements could be transformed into \textit{shares} of a synthetic dataset $D_{priv}$ according the following rules for univariate and multivariate statistics.
% Elements in this case could be attributes of a request for data in the law enforcement case e.g., ``request type:urgent/not urgent'' or ``'', or of a patient in a medical study e.g., ``has gene X:yes/no'' or ``has abnormal blood pressure: yes/no''.

In the univariate case, the elements of a record are simply split and shuffled, which allows univariate statistics to be re-computed without the possibility of computing multivariate statistics.
In the multivariate case, the record in $D$ generates a combination of shares that introduce noise into the $D_{priv}$ but preserve the relative counts between elements in $D$ and therefore allow multivariate statistics to be re-computed.

In both cases, shares in $D_{priv}$ can be tagged so that the published statistics and their inputs can be verified.
The privacy loss associated with the verification of the statistics through a public dataset $D_{priv}$ is also mitigated by ensuring $D_{priv}$ cannot be used to reconstruct $D$ or learn more information than what is learnt from the statistics in the first place, and (in the multivariate case) by generating it in a way that guarantees only a small expected privacy loss from having one's data included.

\subsubsection*{Generating $D_{priv}$ for univariate statistics}
%\alex{Not supporting multivariate statistics is good. DPriv allows stats to eb verified but the stats themselves are not necessarily privacy preserving - some restrictions can be necessary.}
Support for univariate statistics, and in particular a restriction to univariate statistics is important in settings where multivariate statistics are not used due to privacy concerns.
This is for example the reason that IPCO reports contain only univariate statistics.
When only univariate statistics are required, it is, therefore, important that $D_{priv}$ should not be useful to compute multivariate statistics.

To handle the univariate case, shares in $D_{priv}$ are obtained by splitting records in $D$ into shares that each have one element only and then shuffling them, as in Algorithm~\ref{alg:univariate}.
Each share is tagged with the element type and a unique share identifier $id_{share}=Hash(\idc\|i)$ (using a secure hash function e.g., SHA-256) derived from the common identifier $\idc$ of the user and the index $i$ of the share.

\begin{algorithm}
\caption{$\mathcal{M}_1:D\mapsto D_{priv}$ for univariate statistics.}
\label{alg:univariate}
\KwIn{$D=\{r_{i}=(r_{i,1},\ldots,r_{i,\vert r_i \vert}) \vert i\in[1,\vert D \vert], r_{i,j}\in\{\Box,\XBox\}^2 \}$}
\KwOut{$D_{priv}=\{shares\in\{\Box,\XBox\}^2\}$}
list $shares=[]$ \\
\For{$i=1,\ldots,\vert D \vert$}{
  \For{$j=1,\ldots,\vert r_i \vert$}{
  $shares = shares + r_{i,j}$
  }
  }
  $shuffle(shares)$\\
  \Return{$shares$}

\end{algorithm}

Splitting up the shares ensures that the original record cannot be reconstructed without knowing the common identifiers used to generate the share identifiers, so the only thing that can be learnt is the individual counts of elements, which were already revealed by the statistics themselves.
The univariate statistics (i.e., counts) will be unchanged as they do not depend on more than one element so they are unaffected by the split of elements and can be verified.
Multivariate statistics on the other hand will not be computable.

\subsubsection*{Generating $D_{priv}$ for multivariate statistics}
In cases where multivariate statistics are needed (e.g., medical studies), records in $D$ are split into $n$ shares that have as many elements as the records.
Our approach is inspired by Rivest's ThreeBallot voting scheme~\cite{threeballot,rivest2007three}, which works by giving voters three ballots (instead of one) and having them fill them according to a set of rules.
We extend this to any odd number of ballots (combinations of shares in $D_{priv}$) where a vote for or against an element corresponds to having an attribute or not.
We show how to use this as a way of generating a synthetic dataset ($D_{priv}$) that can be used to re-compute multivariate statistics originally computed using $D$.

Shares are generated such that the correct value of each element $e$ (i.e., $\XBox\Box$ or $\Box\XBox$) appears $s\in[1,k+1]$ times and the ``false'' value $\overline{e}$ (i.e., $\Box\XBox$ or $\XBox\Box$) appears $s-1$ times. 
The remaining elements in the shares are neutral and take the form $\XBox\XBox$ or $\Box\Box$. % e.g., the third element of the first share in Figure~\ref{fig:multiballot}. 
Algorithm~\ref{alg:gen} summarises this process.
Once the shares are generated, they are tagged as in the univariate case with a share identifier, split up, and shuffled.

\begin{algorithm}
\caption{Generating valid combinations of shares for $D_{priv}$ in the multivariate case.}
\label{alg:gen}
\KwIn{$n=2k+1, k\in \mathbb{N}$}
\KwOut{Valid combinations of $n=2k+1$ shares for elements $\Box\XBox$ and $\XBox\Box$}
\For{$e$ in $[\XBox\Box,\Box\XBox]$}{
  list $shares_e=[]$ \\
  \For{$s=0,\ldots,k-1$}{
  tuple $shares=(e)$\\
    \For{$i=0,\ldots,s-1$}{
      $shares=shares+(e)+(\overline{e})$\\
    }
    \While{$length(shares)<2k+1$}{
    $shares=shares+(\XBox\XBox)+(\Box\Box)$ \\
    }
    $shares_e=shares_e+permutations(shares)$\\
  }
  }
  \Return{$shares_{\SubXBox\Box,n}$, $shares_{\Box\SubXBox,n}$}
\end{algorithm}

\begin{algorithm}
\caption{$\mathcal{M}_n:D\mapsto D_{priv}$ for multivariate statistics.}
\label{alg:multivariate}
\KwIn{$D=\{r_{i}=(r_{i,1},\ldots,r_{i,\vert r_i \vert}) \vert i\in[1,\vert D \vert], r_{i,j}\in\{\Box,\XBox\}^2 \}$ \\
$n=2k+1, k\in \mathbb{N}$ \\ 
$shares_{\SubXBox\Box}$ \\ 
$shares_{\Box\SubXBox}$
}
\KwOut{$D_{priv}=$}
list $shares=[]$ \\
\For{$i=1,\ldots,\vert D \vert$}{
  tuple $shares_i$ \\
  \For{$j=1,\ldots,\vert r_i \vert$}{
  \eIf{$r_{i,j}=\XBox\Box$}{$\_ \leftarrow shares_{\SubXBox\Box}$ \\ $shares_i = shares_i + \_$ }
  {$\_ \leftarrow shares_{\XBox\SubXBox}$ \\ $shares_i = shares_i + \_$ }
  }
  }
  $shuffle(shares)$\\
  \Return{$shares$}

\end{algorithm}

% There are many valid combinations of shares for a given record, one of which is picked at random.
The number of valid combinations $B$ of shares, given by Equation~\ref{eq:combinations}, is obtained by considering the number of multiset permutations of the shares for each possible value of $s$, summing over $s$, and multiplying by a factor of $2$ to account for both elements.
% Once the shares are generated, they are tagged as in the univariate, split up, and shuffled.

\begin{equation}\label{eq:combinations}
B=2\sum_{s=1}^{k+1}\frac{(2k+1)!}{s!(s-1)!(k+1-s)!(k+1-s)!}
\end{equation}

As an example we look at the record in Figure~\ref{fig:multiballot}, which corresponds to $[\XBox\Box,\XBox\Box,\Box\XBox,\Box\XBox,\XBox\Box]$.
With $n=3$, an element $\Box\XBox$ can generate a combination of shares $(\Box\XBox,\Box\XBox,\XBox\Box)$ and its $6$ permutations, and $(\Box\XBox,\XBox\XBox,\Box\Box)$ and its $3$ permutations.
Similarly, an element $\XBox\Box$ can generate a combination of shares $(\XBox\Box,\XBox\Box,\Box\XBox)$ and its $6$ permutations, and $(\XBox\Box,\XBox\XBox,\Box\Box)$ and its $3$ permutations.
For each element, a valid combination of shares is picked at random.
In Figure~\ref{fig:multiballot} this results in [$(\XBox\Box,\Box\Box,\Box\Box)$, $(\XBox\Box,\XBox\Box,\Box\XBox)$, $(\XBox\XBox,\Box\XBox,\Box\Box)$, $(\XBox\Box,\Box\XBox,\Box\XBox)$, $(\Box\XBox,\XBox\Box,\XBox\Box)$].
Split up in $D_{priv}$, these will look like three records of the form $[\XBox\Box,\XBox\Box,\XBox\XBox,\XBox\Box, \Box\XBox]$, $[\Box\Box,\XBox\Box,\Box\XBox,\Box\XBox,\XBox\Box]$, $[\Box\Box,\Box\XBox,\Box\Box,\Box\XBox,\XBox\Box]$.
% given $D$

Valid combinations of shares can be pre-computed, so all that is required given $D$ is to randomly pick combinations of shares for all records.
The shares are therefore picked from a distribution given in Equations~\ref{eq:probsing} and~\ref{eq:probdoub} for each element, which is derived in more detail in Section~\ref{sec:security}, where it is also shown that the process of generating $D_{priv}$ incurs only a small expected privacy loss.

% Algorithm~\ref{alg:multiballot} summarises this.
\begin{align}
&S_{\SubXBox\Box}=S_{\Box\SubXBox}=\sum_{s=1}^{k+1}(2s-1)\frac{(2k+1)!}{s!(s-1)!(k+1-s)!(k+1-s)!} \label{eq:numsing} \\
&S_{\SubXBox\SubXBox}=S_{\Box\Box}=2\sum_{s=1}^{k+1}(k+1-s)\frac{(2k+1)!}{s!(s-1)!(k+1-s)!(k+1-s)!} \label{eq:numdoub} \\
&\Pr(\XBox\Box)=\Pr(\Box\XBox)=\frac{S_{\SubXBox\Box}}{(2k+1)B} \label{eq:probsing} \\
&\Pr(\XBox\XBox)=\Pr(\Box\Box)=\frac{S_{\SubXBox\SubXBox}}{(2k+1)B} \label{eq:probdoub}
\end{align}

The level of privacy depends on the original distribution of elements in $D$ and the number of shares generated (i.e. the parameter $k$). This means that the level of privacy can be tuned by varying $k$ (see Section~\ref{sec:security}).
Reconstructing a record in $D$ will also be highly improbable as given even all but one of the shares generated from a specific record, finding the correct last share will be improbable as no link will be revealed by the share identifiers.
Verifying multivariate statistics will be possible (as we will see next), as well as their inputs by using the share identifiers.
% \begin{algorithm}
% \SetAlgoLined
% \SetKwInOut{Input}{input}\SetKwInOut{Output}{output}
% \Input{$D=\{record_1,\dots,record_{\vert D \vert}\}$, k}
% \Output{$D_{priv}=\{share_1,\dots,share_{(2k+1)\vert D\vert}\}$}
% \For{$r$ in $D$}{
%   list $s=[]$
%   \For{$e$ in $r$}{
%     pick $shares$ from $shares_{e,k}$ \\
%     $s.append(shares)$
%     }
%   }
% \caption{Generation of $D_{priv}$ from $D$ with MultiBallot}
% \label{alg:multiballot}
% \end{algorithm}

\subsubsection*{Verifying statistics with MultiBallot}

Verifying univariate statistics is straightforward because this just involves counting the number of occurrences of an element having values $0$ or $1$ in $D_{priv}$, which will be the same as the number of occurrences in $D$.
Because shares in $D_{priv}$ are tagged with an element type, the total number of shares that correspond to a specific element is also the same as the number of records in $D$ that include that element.

The multivariate case is a bit more involved and we show how to do it explicitly for association rule mining, although a similar approach could be used for other ways of computing multivariate statistics.
This allows the results of an analysis of $D$ to be estimated from $D_{priv}$ by computing a matrix populated with the expected counts of shares in $D_{priv}$, which translates between $D$ and $D_{priv}$.

Association rule mining~\cite{agrawal1993mining} is one of the most commonly used approaches to identify \emph{if-then} rules and relationships between variables in large datasets.
Given an element set $E$ of binary elements of a record and a dataset $D$ of records containing elements that form a subset of $E$, rules such as $\epsilon \Rightarrow \epsilon^{\prime}$ where $\epsilon,\epsilon^{\prime}\subseteq E$ are used to find interesting relationships between variables e.g., linking a set of genes with a particular disease.
Two measures are commonly used to select interesting rules: 
\textit{support} and \textit{confidence}.

Support, defined in Equation~\ref{eq:support}, indicates how frequently a subset of elements appears in the dataset i.e., the proportion of records $R\in\delta$ (where $\delta\subseteq D$) that contain a subset of elements $\epsilon\in E$.
% The support of a rule $\epsilon\Rightarrow\epsilon^{\prime}$, is simply the support of the joint element sets i.e., $supp(\epsilon\Rightarrow\epsilon^{\prime})=supp(\epsilon\cup\epsilon^{\prime})$.
Confidence, defined in Equation~\ref{eq:confidence}, indicates how often a rule is found to be true.
Given a rule $\epsilon\Rightarrow\epsilon^{\prime}$, it is defined using the support of the rule $\epsilon\Rightarrow\epsilon^{\prime}$ and the support of $\epsilon$.
\begin{align}
&supp(\epsilon)=\frac{\abs{\{R\in\delta:\epsilon\in R\}}}{\abs{\delta}} \label{eq:support} \\
&conf(\epsilon\Rightarrow\epsilon^{\prime})=\frac{supp(\epsilon\Rightarrow\epsilon^{\prime})}{supp(\epsilon)} \label{eq:confidence}
\end{align}
% \begin{equation}\label{eq:support}
% supp(\epsilon)=\frac{\abs{\{R\in\delta:\epsilon\in R\}}}{\abs{\delta}}
% \end{equation}

% Confidence, defined in equation~\ref{eq:confidence}, indicates how often a rule is found to be true.
% Given a rule $\epsilon\Rightarrow\epsilon^{\prime}$, it is defined using the support of the rule $\epsilon\Rightarrow\epsilon^{\prime}$ and the support of $\epsilon$.
% \begin{equation}\label{eq:confidence}
% conf(\epsilon\Rightarrow\epsilon^{\prime})=\frac{supp(\epsilon\Rightarrow\epsilon^{\prime})}{supp(\epsilon)}
% \end{equation}
% Computing these values on $D$ is straightforward but some pre-processing is needed to extract them from $D_{priv}$, which contains elements $\overline{e}$, so not all of the observed values for $\epsilon$ and $\epsilon^{\prime}$ in $D_{priv}$ match those of the original records.

Our goal is to estimate the true counts of $\epsilon$, $\epsilon^{\prime}$ and $\epsilon\cup\epsilon^{\prime}$ in $D$ based on observations from $D_{priv}$, which also contains noise in the form of elements $\overline{e}$.
Computing the support and confidence measures defined above is then straightforward. This process is often referred to as \textit{support recovery}. 
For simplicity, we represent both the original records and the shares as bitstrings.
For example, the record and shares in Figure~\ref{fig:multiballot} can be represented as $[10,10,01,01,10]$ and $[(10,10,11,10,01),(00,10,01,01,10),(11,01,00,01,10)]$.
This is the same as the previous notation with $\XBox=1$ and $\Box=0$.

We define $o_D$ and $o_{D_{priv}}$, which contain the number of occurrences of all possible bitstring permutations in $D$ and $D_{priv}$ in Equation~\ref{eq:o}.
(The number of occurrences may be $0$ for some permutations.)

\begin{equation}\label{eq:o}
o_D,\ o_{priv}=\begin{bmatrix} \#[0] \\ \vdots \\ \#[2^t-1] \\ \end{bmatrix}_{D,D_{priv}}
\end{equation}

We also define $M$ in Equation~\ref{eq:m}.
This matrix stores the expected bitstring occurrences $\mathbb{E}(\#[s])$ of any bitstring $s\in[0,2^t-1]$ in $D_{priv}$ (i.e., $\mathbb{E}[o_{priv}]$) for all possible bitstring permutations and a fixed number of bits $t$.
The value $\mathbb{E}(\#[s])$ is obtained from the distribution of shares given in Equations~\ref{eq:probsing} and~\ref{eq:probdoub}.

\begin{equation}\label{eq:m}
M = \begin{bmatrix} \mathbb{E}[\#[0]]_0 & \dots & \mathbb{E}[\#[0]]_{2^t-1} \\ \threevdots & \threevdots & \threevdots \\  \mathbb{E}[\#[2^t-1]]_0 & \dots & \mathbb{E}[\#[2^t-1]]_{2^t-1} \end{bmatrix}
\end{equation}

A relation between $o_{priv}$, $M$, and $o_{D}$, can be established from the fact that for each record in $D$ its elements contribute an expected amount of each element in $D_{priv}$.
Therefore, the number of occurrences of any given bitstring $s$ in $o_{priv}$ is, on expectation, the sum of the expected amount of that bitstring due to a bitstrings in $D$ times the number of times these bitstrings (denoted $\delta$) occurred in D, as in Equation~\ref{eq:exp}.
Thus, we have that $\mathbb{E}[o_{priv}]$ is simply the result of multiplying $M$ with $o_{D}$, as in Equation~\ref{eq:od}.

\begin{equation}\label{eq:exp}
\mathbb{E}[\#[s]_{D_{priv}}] = \sum_{\delta=0}^{2^t-1} \mathbb{E}[\#[s]]_\delta \cdot \#[\delta]_D
\end{equation}

\begin{equation}
\mathbb{E}[o_{priv}] = M \cdot o_D\label{eq:od} 
\end{equation}

% As the number of records in $D$ grows, it follows from the law of large numbers that $o_{priv}$ will get converge to $\mathbb{E}[o_{priv}]$.
% From Chebyshev's inequality (Equation~\ref{eq:chebyshev}) we can be certain that the probability that $o_{priv}$ (or its entries) differs from $\mathbb{E}[o_{priv}]$ by more than $k$ standard deviations $\sigma$ is less than $\frac{1}{k^2}$.
% In this case, the standard deviation is also small as 

% \begin{equation}\label{eq:chebyshev}
% \Pr [\vert o_{priv} - \mathbb{E}[o_{priv}]\vert \geq k\sigma] \leq \frac{1}{k^2}
% \end{equation}

For the public to verify statistics using values in $o_{priv}$ obtained from $D_{priv}$, the aim is to reverse this process and infer the counts in $o_{D}$.
Alongside $D_{priv}$, $M$ can also be safely released as it does not contain any private information.
Computing its inverse $M^-1$, we can therefore estimate $o_D$ by multiplying $o_{priv}$ with $M^-1$, as in Equation~\ref{eq:opriv}.\footnote{If $M$ is not invertible, it can be made invertible with a change that would not significantly affect the results.}

\begin{equation}\label{eq:opriv}
o_{D}\approx M^{-1} \cdot o_{priv} 
\end{equation}

Based on the inferred value of ${o}_{D}$, the support and confidence measures for any element sets $\epsilon$, $\epsilon^{\prime}$ can then be computed in the usual way.
This allows statistics to be accurately verified, as we show in the evaluation of our implementation in Section~\ref{sec:implementation}.

$D_{priv}$ is used only for verification of the reported statistics, leaving plenty of room for minimizing its information content.
If $D$ is composed of records with a large number of elements, but only a few of these have interesting relations that are relevant in the published statistics, then only these need to be included in $D_{priv}$.
This can significantly reduce the size of $D_{priv}$ compared to $D$.

Our technique can in certain cases support statistics involving continuous variables. 
During the rule mining phase, the researcher may need to examine the exact values (e.g., blood pressure) but once a relevant threshold is identified, all the values can be expressed as larger or smaller than that threshold (e.g., blood pressure over 140/90mmHg).
This practice is common in machine learning algorithms e.g., C4.5 (an extension of ID3~\cite{quinlan1986induction}) builds decision trees from sets of data samples containing both continuous and discrete attributes.
% For those that are continuous, it defines a threshold that maximizes the information gain and then splits the samples to those with attribute value lower than the threshold and those with equal or larger value~\cite{quinlan2014c4}.
Alternatively, continuous variables can be split into multiple binary elements.

%%%%%%%
\section{Operating \sysname}\label{sec:system}
\begin{figure}[t]
\centerline{\includegraphics[width=\linewidth]{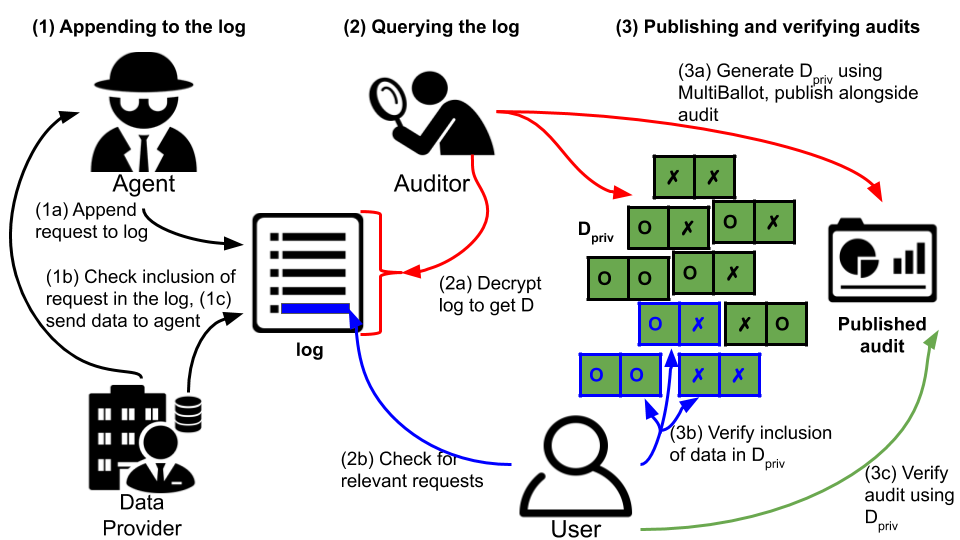}}
  \caption{The three stages in the operation of \sysname. Red, blue and green boxes indicate information available to auditors, users, and the public.
  Similarly, red, blue and green arrows indicate operations that require being an auditor, a user, or anybody.}
  \label{fig:flow}
\end{figure}

\sysname involves three stages that are illustrated in Figure~\ref{fig:flow}: appending requests to the log, querying the log for audits, and publishing and verifying audits.
In this section, we describe each stage and argue that \sysname achieves its transparency and privacy goals.

% We rely on the transparency overlay as a way of providing a tamper evident key-value store that will log requests for data, published audits, and any other relevant events.
% The log entries are tagged such that they can be queried by auditors and users based on their common identifiers.
% Auditors can then release their privacy preserving statistical reports to users that can verify them using MultiBallot.

\subsection{Appending to the log}
As part of $request$, agents append requests to the log as values tied to the relevant common identifiers.
If the value is temporarily sensitive then a cryptographic commitment can be used to ensure that correct logging can be verified after the fact.

Our transparency goals require that the requests be auditable by the parties they pertain to i.e., users performing $\ucheck$ and auditors performing $\audit$, without relying on other parties.
Our privacy goals also require that the private contents of the requests are not visible to any other parties, and that information cannot be inferred about the requests by linking them with other requests, users, agents, or data providers.
This is assured by encrypting the log entries so that only auditors and relevant users can decrypt them, and using unlinkable common identifiers.

Once a request is appended to the log it can be answered by a data provider.
A user acting as a data provider that is relying on a data broker to answer requests for their data could check if the data broker was misrepresenting their preferences by checking access' to their data themselves, or simply receiving notifications for the requests appended to the log that the data broker accepts.

\subsection{Querying the log}
Once requests are logged, users and auditors can verify that the log servers are not malicious by performing $\odetect$, then perform $\audit$ and $\ucheck$ as required.
% This fulfils part of our transparency goals, which also require that users and auditors have access to untampered data.
Both users and auditors can assure themselves that the information obtained from the log is correct due to the availability and integrity properties of the log, and audit the entries of the log.
Auditors perform their task over the entire log, or a subset of the log.
Users look only for specific requests by iterating over their common identifiers until no request is found to determine the possible requests relevant to them.

Users that do not wish to take on this task can choose to outsource it to a data broker.
% These parties act as intermediaries between agents and users that would otherwise perform $\uprovide$, allowing users to act as data providers if they are willing to, for example, participate in a study.
A downside of this is that the data broker must then be trusted with the private identifiers tied to requests that the user positively answers.
No other trust is required as \sysname allows users to check the activity of their broker, which can be logged and audited under the same guarantees as other log entries.

\subsection{Publishing and verifying audits}
Auditors perform $\publish$ to release the statistics computed as a result of their audit.
Examples of what might be published are the statistics provided by the IPCO in its annual report~\cite{ipco2018} (e.g., the number of urgent requests) or the results of a medical study (e.g., the association of some attribute with a disease).

Our transparency goals require that these statistics be verifiable, but for operational and privacy reasons the original data used to compute statistics cannot be published.
Instead, the synthetic dataset $D_{priv}$ generated by MultiBallot (or its hash) can be published on the log and used to verify the statistics by users performing $\monitor$, as we have shown in Section~\ref{subsec:multiballot}.
Users whose data was used to compute the statistics can verify the inclusion of their data in $D_{priv}$ as part of $\monitor$.

Assuming all users will take on the burden of verifying statistics is unrealistic, but the system does not require them to do so.
Users that wish to check for access to their data can do so regardless of others.
They can also verify published statistics even if their data was not used in the computation.
Verifying the integrity of a dataset benefits from more users doing so, but a limited amount of users doing so will already be beneficial.
Others could rely on data brokers, which would be acting in a way similar to organisations that currently perform Freedom of Information requests.

\section{Achieving Transparency And Privacy Goals}\label{sec:security}
% We now discuss the security of \sysname with respect to the transparency and privacy goals stated in Section~\ref{sec:threat}.

% \subsection{Transparency}
% In Section~\ref{sec:transparencygoals}, we defined transparency for \sysname as enabling audits by auditors and users who can be assured of the integrity of the information they audit, and allowing users to verify published audits, without requiring interaction with anything but \sysname.
% Our arguments for transparency therefore cover the availability and integrity of the information accessed for audits, the verifiability of published audits, and the need to rely only on VAMS itself rather than other parties.

\subsection{Goal $T1$: log availability}\label{arg:t1}
We have assumed that agents and data providers do not collude so requests will be logged so availability only requires online log servers.
The remaining threat is then a malicious log server that equivocates, in which case users and auditors could perform $\udetect$ as follows.

In the HLF case, equivocation would result in a fork of the blockchain.
Both the main chain and the forked chain would be visible, so equivocation can be detected.

In the Trillian case, a log server that equivocates would have to produce signed tree heads and Merkle consistency proofs for the alternative Merkle trees.
Different Merkle consistency proofs leading from the same Merkle tree generate different views of the log, but these differing logs can no longer accept the same Merkle consistency proofs to extend the logs because the leaves are different.
As tree heads are signed by the log server, two inconsistent tree heads can be used as evidence to implicate the log server~\cite{dowling2016secure,chase2016transparency}.

\subsection{Goal $T2$: log integrity}\label{arg:t2}
% We now argue that given the availability of the information on the log, users and auditors can also be assured of its integrity.
% Again, the completeness properties of HLF and Trillian will guarantee this.
This argument is based on the fact that updates to the key-value store are recorded on an append-only blockchain (for HLF) or a verifiable log (for Trillian), resulting in \sysname's log being tamper-evident.

(\textit{HLF case}) We rely on the underlying blockchain that records state updates.
Auditors can use the key history function to obtain the state updates that have modified the value of a key.
If they do not trust the integrity of that function (the code for which is public), they can replay the blockchain's transactions to detect a party's misbehaviour as they will have signed the relevant transactions.

(\textit{Trillian case}) We rely on the underlying verifiable log i.e., the Merkle tree and the Merkle consistency proofs that give the append-only property of the tree.
If a malicious party has tried to tamper with requests, they will have to update a request value, which will appear in the append-only log.
If the log server produces a new tree head for a tree that modifies requests in the tree associated with the previous tree head, it will be evident as there cannot be a Merkle consistency proof between the two trees.
Similarly, if a leaf of an existing tree is removed, the Merkle root of the tree will no longer match the leaves.

Auditors can then perform $\audit$ by querying the state of the ledger or log-backed map containing the requests (which are encrypted under their public keys) and performing their analysis.
Requests that cannot be decrypted can be classed as invalid and reported.
The same argument can be used for users performing $\ucheck$.

\subsection{Goal $T3$: verifiability of inputs to audits}\label{arg:t3}
In the case of a user performing $\monitor$, we again have that the user has a correct and complete view of the log by following through the arguments previously presented.
A malicious auditor could nonetheless perform $\publish$ maliciously, publishing incorrect statistics or the wrong dataset, but this would be detected by a user performing $\monitor$.
A user that was included in the used dataset $D$ used can check the integrity of the transformed dataset $D_{priv}$, identifying their shares using the share identifier derived from their common identifiers and checking that they reconstruct their original record.
% Once the integrity of the data is confirmed, any other user can replicate the computations of the analysis and compare their results with those published, detecting any false statistics.

% Finally, if malicious auditor does not generate shares randomly so that $D_{priv}$ might reveal information, users performing $\monitor$ can handle this by checking that the distribution of shares in $D_{priv}$ is as expected.

\subsection{Goal $T4$: verifiability of published audits}\label{arg:t4}
Using $D_{priv}$, any user can compute the same statistics that are contained in the published audits in the way that was described in Section~\ref{subsec:multiballot}.
If the results are acceptably close then they can conclude that the statistics computed on $D$ were correctly computed.

\subsection{Goal $T5$: transparency of the system}\label{arg:t5}
All the information that auditors require is by definition the information that is logged, which they can access with access only to \sysname, and without interaction with any other party.
For users, the information relevant to themselves will be accessible by finding and decrypting the records relevant to them, which does not require the help of any other party, and audits must be made available by the auditors, but a hash of $D_{priv}$ on the log can assert the integrity of $D_{priv}$.
Verifying the statistics from $D_{priv}$ and the inclusion of their data does not require any interaction either.

% \subsection{Privacy}
% We argue that the privacy of the parties in \sysname is protected at each step of \sysname. Because confidential information in the records is encrypted, we focus on the unlinkability of log entries between themselves, and to users, agents or data providers, and the differential privacy of MultiBallot's synthetic data generation.

\subsection{Goal $P1$: The log itself does not reveal any sensitive information}\label{arg:p1}
% We rely on SHA-256, which we take to be a secure cryptographic hash function.
% This implies pre-image resistance, which means it will not be possible to recover the inputs $\ida$ and $\idp$ from common identifiers, and collision resistance, which ensures each identifier is unique.
The values of the log entries are encrypted so that no party can gain any information from these unless they have the decryption key controlled by either a relevant auditor or user.
Identifying related log entries could reveal sensitive information but log entries are unlinkable.
It is not possible to link either common identifiers (outputs of a hash function such as SHA-256) or the values of log entries (outputs of a secure encryption scheme) together.

\subsection{Goal $P2$: verifying an audit is privacy preserving}\label{arg:p3}
Verifying an audit involves verifying known inputs (i.e., privately known records in $D$) and the public outputs (i.e., statistics computed on records that are released).
Verifying the inputs only involves checking that known shares in $D_{priv}$ reconstruct a known record in $D$.
No privacy loss can occur by doing this because the record is, by assumption, already known.
We, therefore, focus on arguing that the access to $D_{priv}$, which is necessary to verify the statistics, does not lead to a greater privacy loss than the release of the statistics themselves.

The privacy risk associated with the release of $D_{priv}$ comes in three forms.

First, the share identifiers used by users to verify the inputs to the statistics could reveal links between the shares, or the common identifiers used as their inputs.
For this, we rely again on the security of the hash function used to generate the share identifiers.
Taking SHA-256 as providing sufficiently random outputs, it will not reveal links between the shares, and taking it as pre-image resistant it will not reveal the common identifiers used as input.

Second, $D_{priv}$ itself may leak sensitive information, allowing a record to be reconstructed or allowing the presence of a record to be inferred more than already possible from the publicly released statistics.

Third, $D_{priv}$ could be used to compute not only the statistics released through the published audit but also other statistics that were not intended to be released.

To verify univariate statistics, $D_{priv}$ needs only to contain single element shares so $D_{priv}$ can only be used to compute univariate statistics.
Moreover, if some elements of the records in $D$ were considered too sensitive to publish statistics about, they can simply be excluded from $D_{priv}$ without affecting the ability of users to compute the statistics that were published.
This means that only the published statistics and their inputs can be verified, so there is no risk of privacy loss from allowing the statistics to be verified.

% \alex{Address point from CCS review about one sided DP~\cite{kotsogiannis2020one}, from rebuttal: VAMS is intended to allow any published statistics to be verified, but it does not determine which statistics can be published (that is the responsibility of auditors). Thus, the idea here is that if a record in D (say a database of physical attributes) contains an element (say, hair colour) that is not a part of any of the published statistics (which pertain only to height and hair length) then there is no need to include hair colour in Dpriv as it will not be needed to re-compute the statistics about height and hair length. It is not a subset of records that are suppressed (this would interfere with the ability to verify the inclusion of one’s data) but the same element across all records - as it does not influence the published statistics it may as well have never been recorded.}

In the case of multivariate statistics, we rely on the fact that generating $D_{priv}$ incurs only a small expected privacy loss (Theorem~\ref{thm:edp}) and that, given $D_{priv}$, it is not possible to reconstruct records on the log (Theorem~\ref{thm:rec}).
This ensures that publishing $D_{priv}$ does not enable an adversary to infer whether or not a certain log entry was in $D$ and that given some information about a record in $D$ (i.e., some shares from that record), the remaining shares cannot be identified using $D_{priv}$. 

\subsubsection{Bounds on ballot reconstruction attacks}

\begin{theorem}\label{thm:rec}
The probability that an adversary who knows $\alpha\in[1,2k]$ shares of a ballot can reconstruct the entire ballot is $\Pr(Reconstruct))=(1-\Pr(Valid)^e)^{\binom{(2k+1)r-\alpha}{2k+1-\alpha}-1}$.
\end{theorem}

\begin{proof}
Each element of a share can take the form of a \textit{single} i.e., $\XBox\Box$ or $\Box\XBox$, or a \textit{double} i.e., $\XBox\XBox$ or $\Box\Box$.
Initially, we restrict ourselves to ballots of one element, so a share simply corresponds to an element.
Given a ballot of $n=2k+1$ elements, it must contain $s\in[1,k+1]$ singles that correspond to the record, and thus $s-1$ copies of the other single.
The rest of the elements are filled up using an equal amount of each double i.e., $k+1-s$ of each.
The number of permutations, denoted $P(s)$, of a ballot with $s$ singles corresponding to the record using the standard formula for multiset permutations, which takes into account repeated elements in a ballot, is given in Equation~\ref{eq:multiset}.

\begin{equation}\label{eq:multiset}
P(s)=\frac{(2k+1)!}{s!(s-1)!(k+1-s)!(k+1-s)!}
\end{equation}

To compute the total number of possible ballots $B$, given in Equation~\ref{eq:numballots}, we just sum over $s$ to add up ballots corresponding to each number of singles matching the record and multiply by a factor $2$ as ballots are symmetric under an interchange of singles.

\begin{equation}\label{eq:numballots}
B=2\sum_{s=1}^{k+1}P(s)
\end{equation}

This result can be used to compute the probability distribution of the shares by counting their appearances in the $(2k+1)B$ shares that make up all the possible ballots.
This amounts to taking, for each element, the sum of permutations of a ballot weighted by the number of appearances of that share in the ballot, and taking into account the fact that doubles appear the same amount of times in ballots corresponding to either record, and singles appear either $s$ or $s-1$ times depending on whether they match the record.
We denote the number of $\XBox\Box$, $\Box\XBox$, $\XBox\XBox$ or $\Box\Box$ shares by $S_{\SubXBox\Box}$, $S_{\Box\SubXBox}$, $S_{\SubXBox\SubXBox}$ or $S_{\Box\Box}$, which are given in Equations~\ref{eq:numsingles} and~\ref{eq:numdoubles}.
The probability of each share, given in Equations~\ref{eq:probsingles} and~\ref{eq:probdoubles}, is then obtained by dividing the number of shares for each form by the total number of shares.

\begin{align}
&S_{\SubXBox\Box}=S_{\Box\SubXBox}=\sum_{s=1}^{k+1}(2s-1)P(s) \label{eq:numsingles} \\
&S_{\SubXBox\SubXBox}=S_{\Box\Box}=2\sum_{s=1}^{k+1}(k+1-s)P(s) \label{eq:numdoubles} \\
&\Pr(\XBox\Box)=\Pr(\Box\XBox)=\frac{S_{\SubXBox\Box}}{(2k+1)B} \label{eq:probsingles} \\
&\Pr(\XBox\XBox)=\Pr(\Box\Box)=\frac{S_{\SubXBox\SubXBox}}{(2k+1)B} \label{eq:probdoubles}
\end{align}

With the probability distribution obtained we can obtain the probability $\Pr(Valid)$, given in Equation~\ref{eq:prv}, of the event $V$ that occurs when randomly chosen shares form a valid ballot, by summing over the possible ballots weighted by the probability of each share.
More generally, when shares involve $e$ elements the probability is $\Pr(Valid)^e$.

\begin{equation}\label{eq:prv}
\begin{split}
\Pr(Valid)=2\sum_{s=1}^{k+1}&P(s)\Pr(\XBox\Box)^{s}\Pr(\Box\XBox)^{s-1} \cdot\\
&\Pr(\XBox\XBox)^{k+1-s}\Pr(\Box\Box)^{k+1-s}
\end{split}
\end{equation}

The above is the probability of success for a weak adversary that starts with no prior knowledge and wants only to reconstruct a ballot, regardless of whether it belongs to someone.
An adversary that knows up to $\alpha\in[1,2k]$ shares of a ballot and wishes to figure out the last shares required to reconstruct that ballot chooses $2k+1-\alpha$ other shares from the dataset, giving $\binom{(2k+1)r-\alpha}{2k+1-\alpha}$ possibilities, where $r$ is the number of records from which we subtract 1 as there must be at least one valid ballot.
This gives Equation~\ref{eq:pradv}, which expresses the probability of success $\Pr(Reconstruct)$ of that adversary.

% \alex{Observation: if a an adversary knows only 4 shares of a record, but these something like oo,xx,xo,xo then they have no need to find the fifth as the last share must be a ox share. Therefore even with $n$-Ballot the security may depend on fewer than $n$ shares. (this was also true for the original threeBallot e.g. when given xo,xo then the last share must be ox. Did we factor this in properly/explicitly enough?)}

\begin{equation}\label{eq:pradv}
\Pr(Reconstruct))=\left( 1-\Pr \left( Valid \right) ^e \right)^{\binom{\left( 2k+1 \right) r-\alpha}{2k+1-\alpha}-1}
\end{equation}
\end{proof}

Table~\ref{table:bounds} gives an upper bound on elements that can be included in shares while maintaining a probability of a reconstruction attack under $0.01$ when an adversary knows one share or all but one share.
Different bounds can be chosen depending on the acceptable probability of a reconstruction. 
In practice, only a few elements may be relevant to the results of an audit or study, and only those need to be published for the relevant statistics to be publicly verifiable.

It is also important to note that we have modelled an attacker who completes a partial ballot by picking random shares in $D_{priv}$. 
In reality, an attacker may of course have better chances of reconstructing a ballot by inferring the remaining shares, particularly if they already know most of the ballot's shares, but this is done regardless of the availability of $D_{priv}$.

Finally, we have assumed statistical independence between elements, which may not always be true.
ThreeBallot with correlated ballots was studied by Strauss~\cite{strauss2006critical} who showed that even heavily correlated elements had only a minor effect on the security of the scheme.

\begin{table}[t]
\begin{center}
\caption{Upper bounds on the number of elements in 3Ballot and 5Ballot shares such that the probability of a successful reconstruction is less than $0.01\%$. The numbers in brackets next to the scheme indicate the number of shares known to the adversary and the numbers in brackets next to the number of elements indicate the probability of success.}
\label{table:bounds}
\resizebox{\linewidth}{!} {%
\begin{tabular}{l l l l l l }
\thead{Scheme} & \thead{10 users} & \thead{100 users} & \thead{1000 users} & \thead{10\ 000 users}\\ 
\midrule
3Ballot (1) & 3 $(3\cdot10^{-5})$ & 6 $(5\cdot10^{-13})$ & 10 $(6\cdot10^{-10})$ & 14 $(1\cdot10^{-7})$ \\
3Ballot (2) & 1 $(8\cdot10^{-5})$ & 2 $(2\cdot10^{-12})$ & 4 $(2\cdot10^{-10})$ & 6 $(5\cdot10^{-9})$ \\
5Ballot (1) & 6 $(4\cdot10^{-6})$ & 11 $(5\cdot10^{-20})$ & 17 $(3\cdot10^{-12})$ & 23 $(2\cdot10^{-13})$\\
5Ballot (4) & 1 $(5\cdot10^{-5})$ & 2 $(3\cdot10^{-9})$ & 3 $(2\cdot10^{-17})$ & 5 $(3\cdot10^{-7})$ \\
\end{tabular}
}
\end{center}
\end{table}

\subsubsection{Bounds on the expected privacy loss from membership of $D_{priv}$}
% \alex{DP->expected privacy loss, then do a discussion of DP.}
% In the next theorem, we derive an upper bound on the expected privacy from membership of $D_{priv}$.
% This means that the effect that a user's record in $D$ has on the version of $D_{priv}$ that results from the use of MultiBallot is, on expectation, hard to observe.
% Therefore, the user's participation in $D$, which is then processed with MultiBallot to produce $D_{Priv}$, does not incur a significant expected loss of privacy.

To quantify the loss of privacy from membership of $D$ and, therefore, the published $D_{priv}$, we consider the privacy loss variable $\mathcal{L}_{M(D), M(D^\prime)}^{\theta}$, defined in Equation~\ref{eq:privloss}.
This variable quantifies the privacy loss incurred by observing an output $\theta$ of the mechanism $M$, based on how much more (or less) likely that output is when $M$ takes $D$ as input rather than $D^\prime$.

\begin{equation}\label{eq:privloss}
\mathcal{L}_{M(D), M(D^\prime)}^{\theta} = \ln\left(\frac{\Pr[\mathcal{M}(D)=\theta]}{\Pr[\mathcal{M}(D^\prime)=\theta]}\right)
\end{equation}

If $M$ satisfied the definition of differential privacy, this would be equivalent to saying that the privacy loss variable would be bounded~\cite{dpbook}.
MultiBallot, however, cannot satisfy differential privacy because the share counts (which would define $D_{priv}$) that result from running $M$ on $D$ or $D^\prime$ cannot ever match unless $D=D^\prime$.
This is because the share counts of ballots generated from different elements cannot be equal.

We, therefore, define in Definition~\ref{def:edp} a relaxed alternative to differential privacy, replacing the distribution over outputs with an expected output.
As we will show in Theorem~\ref{thm:edp}, MultiBallot satisfies this definition such that given two datasets of the same size, $D$ and $D^\prime$, differing in one entry, the expected outputs of $M_n$ running on either database remain close.

\begin{definition}[$\zeta$-expected privacy loss]\label{def:edp}
A randomised algorithm $M$ with domain $\mathbb{N}^{\left| \chi \right|}$ has a bounded expected privacy loss if given two input databases $D,\ D^\prime \in \mathbb{N}^{\left| \chi \right|}$ where $D$ and $D^\prime$ differ only in one element, there exists $\zeta \in \mathbb{R}$ such that
% \begin{equation}\label{eq:ediffp}
% \zeta\geq\ln\left(\frac{\Pr[\mathcal{M}_n(D)=\theta]}{\Pr[\mathcal{M}_n(D^\prime)=\theta]}\right)
% \end{equation} % $$\Pr[\mathcal{A}(D)=\theta]\leq e^\epsilon\Pr[\mathcal{A}(D^\prime)=\theta]$.

\begin{equation}\label{eq:ediffp}
\zeta\geq\ln\left(\frac{\mathbb{E}[M(D)]}{\mathbb{E}[M(D^\prime)]}\right).
\end{equation} 

\end{definition}

As in the case of differential privacy, $\zeta$-expected privacy loss also provides group privacy.

\begin{lemma}[Bounded expected group privacy loss]
Let $D$ and $D^\prime$ be two databases that differ in $e$ elements. If a randomised algorithm $M$ satisfies bounded expected privacy loss, then we have that

\begin{equation}\label{eq:ediffp}
e\zeta\geq\ln\left(\frac{\mathbb{E}[M(D)]}{\mathbb{E}[M(D^\prime)]}\right).
\end{equation} 

\end{lemma}
\begin{proof}
Iterating over Definition~\ref{def:edp}, if $M$ has bounded expected privacy loss then the expected privacy loss due to any single element on the output of $M$ is bounded by $\zeta$.
Thus, the impact of any $e$ elements is bounded by $e\zeta$.
\qed
\end{proof}

We now prove in Theorem~\ref{thm:edp} that Multiballot satisfies this definition and compute values of $zeta$ for different database sizes ($\left| D \right| = 10, 100, 1000, 10000$) and schemes (3Ballot, 5Ballot).

% \alex{Feeling a bit iffy about this - it's looking at the counts rather than the possible databases themselves, although stats are computed based on counts. In particular it seems like in the case where one record has a count of 0 in $D$ and 1 in $D^\prime$ then there should be an impossible event for $D^\prime$ that can occur for $D$, which is to generate a $D_{priv}$ that has zero of the record that must exist in $D^\prime$. This should introduce a $\delta$.}

% \alex{rewrite in terms of epxectation values i.e. some probabilities should be expectations}
\begin{theorem}\label{thm:edp}
The MultiBallot share generation mechanism $M_{n}:D \mapsto D_{priv}$ satisfies $\zeta$-expected privacy loss with $\zeta=\ln \left( \frac{\left\vert\XBox\Box\right\vert_{D_{priv}}}{\left\vert \XBox\Box \right\vert_{D_{priv}}-\sum_{s=1}^{k+1}\Pr \left( s\cdot\XBox\Box \right)}\right)_{\big\vert_{r_{\SubXBox\Box}=0,r_{\Box\SubXBox}=r}}$.
\end{theorem}

\begin{proof}
% \alex{rewrite in terms of expectation values.}
Consider two databases $D$ and $D^\prime$ that contain $r$ single element records and differ in one record.
Without loss of generality, we take $D$ to contain $r^{D}_{\SubXBox\Box}$ records of the form $\XBox\Box$ and $r^{D}_{\Box\SubXBox}$ records of the form $\Box\XBox$ and $D^\prime$ to contain $r^{D^\prime}_{\SubXBox\Box}=r^{D}_{\SubXBox\Box}-1$ records of the form $\XBox\Box$ and $r^{D^\prime}_{\Box\SubXBox}=r^{D}_{\Box\SubXBox}+1$ records of the form $\Box\XBox$.

Our aim is to determine and compare the share counts in $D_{priv}\leftarrow M_n(D)$ and $D^\prime_{priv}\leftarrow M_n(D^\prime)$.
Counting the different shares in $D_{priv}$ amounts to considering the probability that the ballot generated from a record will have $s$ shares of one form.
This is given by the number of ballot permutations for a given $s$ over all possible ballots for that record, as expressed in Equation~\ref{eq:ballothasshare}.
Summing over $s$ gives the expected count for each share of ballots generated from a record, which we denote $\vert share\vert_{record}$.
(The same analysis holds for $D^\prime_{priv}$.)

\begin{align}
&\Pr \left( s\cdot \left( share=record \right) \right)=\frac{2}{B}P \left( S \right) \label{eq:ballothasshare} \\
&\left\vert \Box\XBox \right\vert_{\Box\SubXBox}=\vert\XBox\Box\vert_{\SubXBox\Box}=r^{D}_{\SubXBox\Box}\sum_{s=1}^{k+1}\Pr \left( s\cdot\XBox\Box \right) s \label{eq:1} \\
&\vert\XBox\Box\vert_{\Box\SubXBox}=\vert\Box\XBox\vert_{\SubXBox\Box}=r^{D}_{\SubXBox\Box}\sum_{s=1}^{k+1}\Pr \left( s\cdot\XBox\Box \right) \left( s-1 \right) \label{eq:2} \\
&\vert\XBox\XBox\vert_{\Box\SubXBox}=\vert\XBox\XBox\vert_{\SubXBox\Box}=r^{D}_{\SubXBox\Box}\sum_{s=1}^{k}\Pr \left( s\cdot\XBox\Box \right) \left( k+1-s \right) \label{eq:3} \\
&\vert\XBox\XBox\vert_{\Box\SubXBox}=\vert\Box\Box\vert_{\SubXBox\Box}=r^{D}_{\SubXBox\Box}\sum_{s=1}^{k}\Pr \left( s\cdot\XBox\Box \right) \left( k+1-s \right) \label{eq:4}
\end{align}

Adding the contributions from all the records together gives the total expected share count for each type of share in $D_{priv}$.

\begin{align}
\begin{split}\label{eq:5}
    \left\vert \XBox\Box \right\vert_{D_{priv}} ={}& r^{D}_{\SubXBox\Box}\sum_{s=1}^{k+1}\Pr \left( s\cdot\XBox\Box \right)s\\
         & +r^{D}_{\Box\SubXBox}\sum_{s=1}^{k+1}\Pr \left( s\cdot\Box\XBox \right) \left( s-1 \right)
\end{split}\\
\begin{split}\label{eq:6}
    \left\vert \Box\XBox \right\vert_{D_{priv}} ={}& r^{D}_{\SubXBox\Box}\sum_{s=1}^{k+1}\Pr \left( s\cdot\XBox\Box \right) \left( s-1 \right)\\
         & +r^{D}_{\Box\SubXBox}\sum_{s=1}^{k+1}\Pr \left( s\cdot\Box\XBox \right)s
\end{split}\\
\begin{split}\label{eq:7}
    \left\vert \Box\Box \right\vert_{D_{priv}} ={}& \vert\XBox\XBox\vert =r^{D}_{\SubXBox\Box}\sum_{s=1}^{k}\Pr \left( s\cdot\XBox\Box \right) \left( k+1-s \right) \\
    & \qquad\quad +r^{D}_{\Box\SubXBox}\sum_{s=1}^{k}\Pr \left( s\cdot\Box\XBox \right) \left( k+1-s \right)
    \end{split}
\end{align}

We obtain a similar result for $D^\prime_{priv}$ using the fact that $r^{D^\prime}=r^{D^\prime}_{\SubXBox\Box}-1$.

\begin{align}
\begin{split}
    \left\vert \XBox\Box \right\vert_{D^{\prime}_{priv}} = {}& \left( r^{D}_{\SubXBox\Box}-1 \right) \sum_{s=1}^{k+1} s \cdot\Pr \left( s\cdot\XBox\Box \right) +\\
&\quad \left( r^{D}_{\Box\SubXBox}+1 \right) \sum_{s=1}^{k+1} \left( s-1 \right) \cdot \Pr \left( s\cdot\Box\XBox \right) \\
&=\left\vert \XBox\Box \right\vert_{D_{priv}} - \\ 
&\quad \left( \sum_{s=1}^{k+1} s \cdot \Pr \left( s\cdot\XBox\Box \right)- \sum_{s=1}^{k+1} \left( s-1 \right) \cdot \Pr \left( s\cdot\Box\XBox \right) \right) \\
&=\left\vert \XBox\Box \right\vert_{D_{priv}}-\sum_{s=1}^{k+1}\Pr \left( s\cdot\XBox\Box \right)
\end{split}\\
\begin{split}
\left\vert \Box\XBox \right\vert_{D^{\prime}_{priv}} ={}& \left\vert \Box\XBox \right\vert_{D_{priv}}+\sum_{s=1}^{k+1}\Pr \left( s\cdot\Box\XBox \right)
\end{split}\\
\begin{split}
\left\vert \Box\Box \right\vert_{D^{\prime}_{priv}} ={}& \left\vert \Box\Box \right\vert_{D_{priv}}
\end{split} \\
\begin{split}
\left\vert \XBox\XBox \right\vert_{D^{\prime}_{priv}} ={}& \left\vert \XBox\XBox \right\vert_{D_{priv}}
\end{split}
\end{align}

Given the expected share counts in $D_{priv}$ and $D^\prime_{priv}$ that we have derived, we can now compare them to obtain a bound $\zeta$ on the expected privacy loss.

\begin{equation} \label{eq:epss}
\begin{split}
\zeta ={}& \max_{s\in{shares}} \ln \left( \frac{\left\vert s \right\vert_{D_{priv}}}{\left\vert s \right\vert_{D^\prime_{priv}}} \right) \\ 
=& \max \ln \left( \frac{\left\vert \XBox\Box \right\vert_{D_{priv}}}{\left\vert \XBox\Box \right\vert_{D^\prime_{priv}}} \right) \\ 
% =& \max \ln \left( \frac{\frac{\left\vert \XBox\Box \right\vert_{D_{priv}}}{nr}}{\frac{\left\vert \XBox\Box \right\vert_{D^\prime_{priv}}}{nr}} \right) \\ 
% =& \max \ln \left( \frac{\left\vert \XBox\Box \right\vert_{D_{priv}}}{\left\vert \XBox\Box \right\vert_{D^\prime_{priv}}} \right) \\ 
=& \max \ln \left( \frac{\left\vert \XBox\Box \right\vert_{D_{priv}}}{\left\vert \XBox\Box \right\vert_{D_{priv}}-\sum_{s=1}^{k+1}\Pr \left( s\cdot\XBox\Box \right)} \right) \\ 
=& \ln \left(  \frac{\left\vert \XBox\Box \right\vert_{D_{priv}}}{\left\vert \XBox\Box \right\vert_{D_{priv}}-\sum_{s=1}^{k+1}\Pr \left( s\cdot\XBox\Box \right)} \right)_{\big\vert_{r_{\SubXBox\Box}=0,r_{\Box\SubXBox}=r}}
\end{split}
\end{equation}
\end{proof}

The final result from Equation~\ref{eq:epss} can be easily computed and is given for different values of $\vert D \vert$ in Table~\ref{table:ep}.

\begin{table}[t]
\begin{center}
\caption{Values for the expected privacy loss parameters $\zeta$ and $e\zeta$ for different sizes of $D$. We take values of $e$ equal to the safe number of elements against reconstruction attacks taken from Table~\ref{table:bounds}.}
\label{table:ep}
\resizebox{\linewidth}{!} {%
\begin{tabular}{l l l l l l l}
\thead{Scheme} & $\vert D \vert$ & \thead{$\zeta$} & $\exp(\zeta)$ & \thead{$e\zeta$} & $\exp(e\zeta)$\\ 
\midrule
3Ballot & $10$ & 0.36 & 1.43 & 1.08 ($e=3$) & 2.95 \\
3Ballot & $100$ & 0.03 & 1.03 & 0.18 ($e=6$) & 1.2 \\
3Ballot & $1000$ & 0.003 & 1.003 & 0.03 ($e=10$) & 1.03 \\
3Ballot & $10,000$ & 0.0003 & 1.0003 & 0.0042($e=14$) & 1.0042 \\
\midrule
5Ballot & $10$ & 0.1335 & 1.143 & 0.801($e=6$) & 2.23 \\
5Ballot & $100$ & 0.0126 & 1.0127 & 0.1386 ($e=11$) & 1.149 \\
5Ballot & $1000$ & 0.00125 & 1.00125 & 0.02125 ($e=17$) & 1.0215 \\
5Ballot & $10,000$ & 0.000125 & 1.000125 & 0.002875 ($e=23$) & 1.0029 \\
\end{tabular}
}
\end{center}
\end{table}

%%%%%%%%%
\section{Implementation and Performance}\label{sec:implementation}
We evaluate \sysname by comparing two implementations of the log based on HLF and Trillian (the code for which will be open-sourced after publication) and showing that statistics can be accurately verified with MultiBallot.
Both log implementations are evaluated on very modest (and cheap) Amazon AWS t2.medium instances.\footnote{Each instance has 2 vCPUs and 4GB of memory and is running Ubuntu Linux 16.04 LTS with Go 1.7, docker-ce 17.06, docker-compose 1.18, and Fabric 1.06 installed}

\subsection{Evaluating Hyperledger Fabric and Trillian based logs}
\paragraph*{Hyperledger Fabric implementation}

\begin{figure}[t]
\centerline{\includegraphics[width=\linewidth]{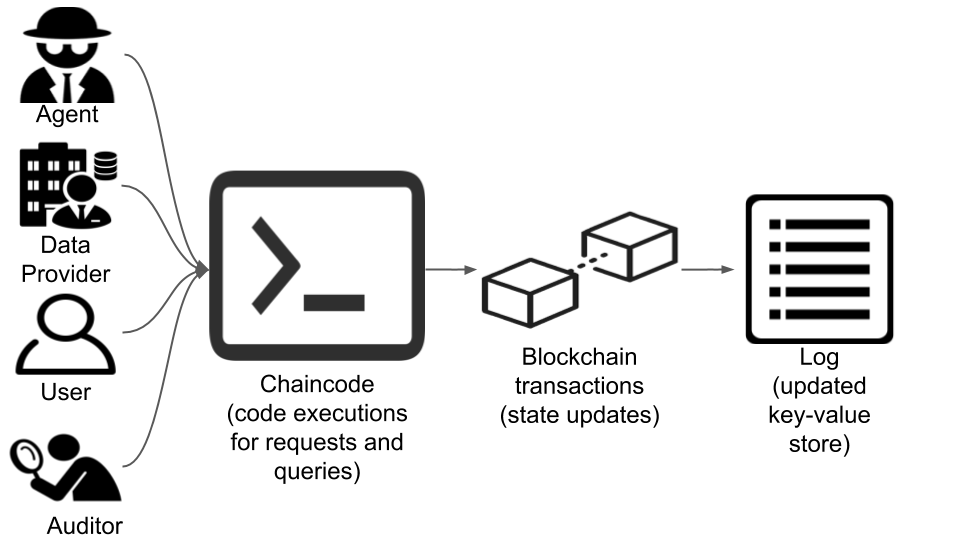}}
  \caption{The HLF-based implementation.}
  \label{fig:hlf}
\end{figure}

For our HLF-based log, we set up a test network of seven machines that represent four peers (an agent, a data provider, a user, and an auditor), an ordering service (an Apache Zookeeper service and a Kafka broker), and a client from which commands are sent.
% The network maintains a key-value store that can be populated by requests tied to common identifiers.
Log entries can be retrieved by querying specific common identifiers, and a key history function is also available to retrieve the state updates (i.e., transactions on the underlying blockchain) of a log entry.
The execution of commands on the HLF network is summarized in Figure~\ref{fig:hlf}.

In this implementation, all peers are connected to one channel and there is one chaincode containing four functions that update the state of the ledger (as part of $\request$), retrieve a range of key values (as part of $\audit$), retrieve values for specific keys (as part of $\ucheck$) and retrieve a key's history (as part of $\audit$ and $\ucheck$).
% to see which blockchain transaction resulted in state updates for a given key.
% The transactions that result in state updates (i.e., change the value of a key) correspond to chaincode invocations, which are recorded on the underlying blockchain.
% Chaincode invocations that only query the log (i.e., do not update the value of a key) are also recorded.

% Peers are members of organizations (one corresponding to each role) and have identities, X.509 public-key certificates, and sign transactions accordingly.
% The public key certificates are issued by the HLF CA, but any organization can specify the CA they wish to issue certificates from, or use another public key infrastructure (PKI) if they wish to do so.
% Signatures are checked as part of the transaction process, as chaincode invocations must be endorsed (signed) by the appropriate parties.
% In our implementation, these are the peers invoking the chaincode.
Thus, auditors or users can check the transactions that updated the value of a key and easily determine the agent responsible for the update, as they will have endorsed (i.e., signed) the transaction.
Endorsement policies can require multiple signatures so they could hold multiple parties accountable e.g., if data providers were considered responsible for accepting invalid requests, they could be required to sign the corresponding request transactions.
An ordering service of specific peers (e.g., auditors) could also be used to detect and flag invalid requests as they are initially processed (and endorsement policies are checked) before committing the requests.
These are not present in our implementation, but give an idea of what improvements may be possible as Hyperledger Fabric undergoes continued development and implements further cryptographic tools.

\paragraph*{Trillian implementation}
% Our second implementation of the log, illustrated in Figure~\ref{fig:trill}, is based on Trillian's verifiable log-backed map.

\begin{figure}[t]
\centerline{\includegraphics[width=\linewidth]{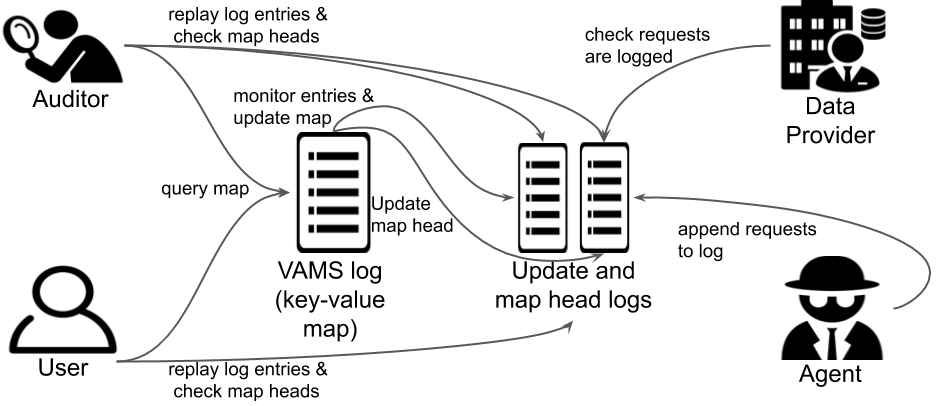}}
  \caption{The Trillian-based implementation.}
  \label{fig:trill}
\end{figure}

% As part of $\request$, agents append signed requests to the log, which data providers can then check.
% Unlike HLF, there is no built-in identity system so the log server service responsible for receiving new requests must check that they are signed.
Our second implementation of the log, illustrated in Figure~\ref{fig:trill}, is based on Trillian's verifiable log-backed map.
The map server (\sysname's log) monitors the log of updates for new entries and updates the map according to the new entries -- common identifiers are used as the map's keys.
It then periodically publishes signed map heads on the second verifiable log, solely responsible for keeping track of published signed map heads.

To perform $\ucheck$, users can query the map to efficiently check their possible common identifier values.
The map will return a Merkle proof of non-inclusion for common identifiers that do not map to requests (i.e. the common identifier maps to 0), or a Merkle proof of inclusion for requests that the common identifiers do map to.
Auditors performing $\audit$ can in turn check that the map is operated correctly by replaying all log entries, verifying that they correspond to the map heads on the second verifiable log.

\paragraph*{Performance}
Table~\ref{table:micro-benchmarks} presents benchmarks for state updates, state retrievals, and the maximal throughput for each system with a batch size of one.
In both cases, the average for each operation is a few dozen milliseconds.
For the HLF system, the results include the time required to create and submit 500 blocks; chaincode execution alone is under $10ms$.
For state retrievals, HLF allows values to be retrieved for a range of keys.
This operation scales linearly with the number of values retrieved and only requires one transaction.
% In the case of state updates and retrievals, the results were obtained by averaging over 500 operations.

\begin{table}[t]
\caption{Micro-benchmarks of basic operations for the Hyperledger Fabric and Trillian based implementations. The maximal throughput values are given for a batch size of 1 in the HLF case and a batch size of 300 in the Trillian case.}
\label{table:micro-benchmarks}
\begin{center}
\begin{tabular}{l c c}
\thead{Measures} & HLF & Trillian  \\
\midrule
State update (average over 500 operations) & 65ms & 35ms \\
Request retrieval (average over 500 operations) & 66ms & 14ms \\
Max throughput & 40 & 102 \\
\end{tabular}
\end{center}
\end{table}

\begin{figure}[t]
\centerline{\includegraphics[width=\linewidth]{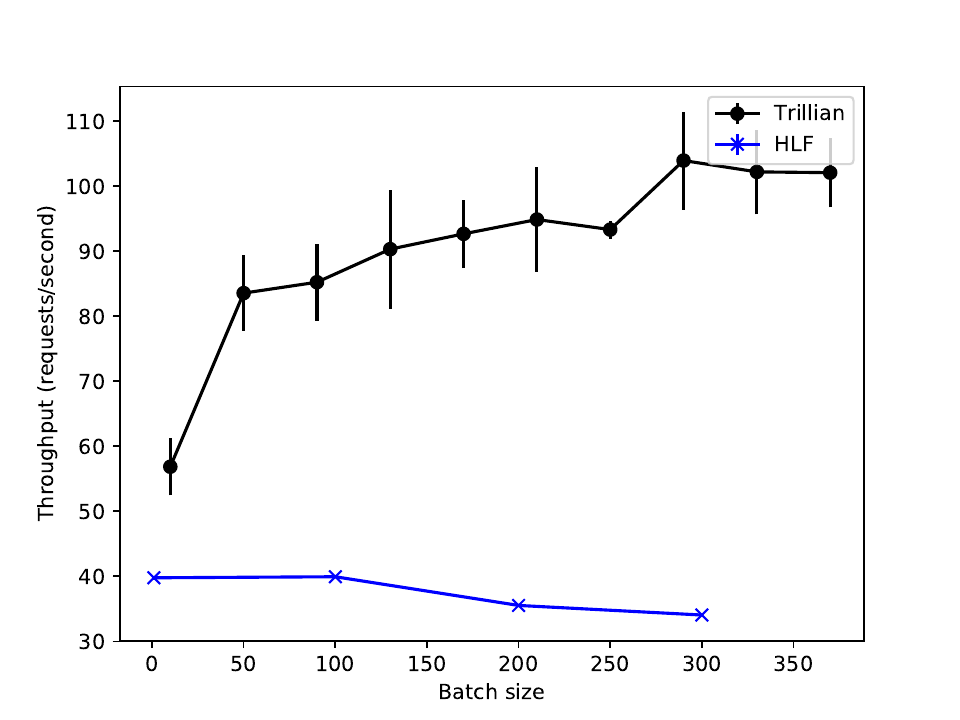}}
  \caption{Throughput of both logs for different batch sizes.}
  \label{fig:trillian-throughput}
\end{figure}

Table~\ref{table:micro-benchmarks} also includes the maximal throughput, which is 40 requests per second for the HLF system and 102 requests per second for the Trillian system.
Figure~\ref{fig:trillian-throughput} shows the throughput for different batch sizes.
For the HLF-based log, the highest throughput is observed for smaller batch sizes.
The bottleneck is simply the client sending requests.
% Throughput then lowers slightly as batch size increases.
For the Trillian-based log, the batch size determines how many items at a time the map servers retrieve from the log to update the map's key values, until around batch size 300.
The bottleneck is then the number of keys updated by the map server per second, and throughput levels out.

% There are, however, trade-offs to consider between batch size and throughput.
A greater throughput can be achieved with a larger batch size but having requests appear on the log sooner can be advantageous e.g., in the case of urgent requests.
In that case, a lower batch size is preferable, or a batch time-out that would ensure a request will appear after a time limit if the batch size limit is not reached.
%Thus, a lower batch size may be advantageous to ensure requests appear as soon as possible, particularly for urgent requests.
% A batch timeout can also be used as a compromise, such that a high batch size can be chosen with a guarantee that a request will appear after a time limit if the batch size limit is not reached.

Practically speaking, for example in the case of law enforcement access to communications data, the IPCO reports about $800\,000$ requests for communication data per year in the UK~\cite{ipco2018} or about 1 request every 9 seconds assuming that requests happen during work hours. (There are no equivalent publicly available statistics for other settings.)

A HLF-based log capable of 40 requests per second, placed at the interface for law enforcement (standardized by ETSI TS 103 307~\cite{ETSI}) would be more than sufficient, with an average waiting time of 25 ms assuming Poisson-distributed requests.
For a Trillian-based system with 102 transactions per second, the average waiting time would be 10 ms.

\paragraph*{Trade-offs}\label{sec:tradeoff}
Trillian has a higher throughput as no consensus is required among different nodes to agree on the ordering of transactions, and better user auditability as when a user queries the map server for an $id_c$, the map server returns a Merkle proof of the key and value being included in the map.
The key history function of HLF does not provide a cryptographic proof, so replaying the entire blockchain can be necessary to verify the inclusion of a key and value.
% This could be managed if ``light clients'' were introduced (as in Ethereum).
Users could however outsource this task to a data broker.

HLF supports flexible chaincode policies to determine write access to the log and comes with built-in authentication and PKI services.
However, this means that users must submit queries to audit the log using a pseudonymous identity.
If they used the same identity for multiple queries, their common identifiers could be linked together.
Authentication must be done separately in Trillian.

The two systems also differ in their architecture.
HLF is decentralized (although it is permissioned) whereas Trillian is centralized.
A decentralized approach is appealing because it reduces the trust required in single entities to maintain the log.
In practice, however, there is only one organization that legitimately has reason to write records for a particular business relationship.
Users will mostly only have a single data provider for a service, which may lend itself more towards the centralized approach.

Table~\ref{table:summary} summarizes the features of both implementations.
Ultimately, Trillian is easier to deploy and has less setup than HLF, which requires the setup of a network of multiple nodes to act as peers, and the maintenance of an identity service to allow nodes to interact with the network.
HLF and other blockchain-based approaches may be preferable if an organisation is already using the technology for some other purpose.

\begin{table}[t]
\caption{Summary of supported (full circles) and partially supported (half-circles) features of the HLF and Trillian based logs.}
\label{table:summary}
\begin{center}
\begin{tabular}{l c c }
\thead{Features} & HLF & Trillian \\
\midrule
User privacy & \CIRCLE & \CIRCLE \\
Agent privacy & \LEFTcircle & \CIRCLE \\
Data provider privacy & \LEFTcircle & \CIRCLE \\
Statistical privacy & \CIRCLE & \CIRCLE \\
User auditability & \LEFTcircle & \CIRCLE \\
External auditability & \CIRCLE & \CIRCLE \\
Verifiability & \CIRCLE & \CIRCLE \\
Access control & \CIRCLE & \LEFTcircle \\
\end{tabular}
\end{center}
\end{table}

\subsection{Evaluating the verification of statistics with Multiballot}
The simplest case when verifying statistics is the univariate case.
In this case, the exact counts for each value of every element are preserved, so statistics can be recomputed with 100\% accuracy.
This means that for applications like the IPCO report on law enforcement access to communications data~\cite{ipco2018}, every statistic could be verified using our scheme.
We, therefore, focus on the more complicated case of univariate statistics for the rest of this section.

Our evaluation measures the accuracy of the association rule metrics computed on $D_{priv}$.
For our experiments, we generate multiple synthetic datasets that follow the structure of $D$ described in Section~\ref{sec:system}, with several frequent element sets~\cite{7506659}.
We mine these for association rules using the Apriori algorithm~\cite{agrwal1994first}, identifying frequent elements in the dataset and extending them to larger element sets for as long as the element sets appear frequently enough in the dataset.
We then compute the support and confidence measures on $D_{priv}$ for the previously extracted element sets, and compare those values with the reported values for the same element sets on $D$.
We use the percent error $\%Err$, defined in Equation~\ref{eq:percenterror}, to measure the disparity between statistics computed on $D$ (the ground truth value, $GV$) and $D_{priv}$ (the measured value, $MV$).

\begin{equation}\label{eq:percenterror}
\%Err = \frac{\vert MV - GV\vert}{|GV|}\cdot 100
\end{equation}

% Element sets are commonly extracted both in the interception use case e.g., proportion of urgent requests, analysis of request rejections, errors and recommendations~\cite{iocco2016}, and the healthcare use case e.g., proportions of people registered with diabetes that achieved blood glucose, pressure and cholesterol targets~\cite{nhs2017national}.
We opt to use synthetic datasets to evaluate MultiBallot, by simulating scenarios with a known ground truth rather than relying on sanitized public datasets for which the ground truth is unknown.
We also verify our results using commonly used public datasets, such as the Extended Bakery dataset~\cite{dekhtyarextended} and the T10I4D100K dataset~\cite{flouvat2005thorough}.
In all our experiments (repeated $100$ times) we measure the error for both the support and the confidence metrics.
Because these are identical, however, we only include the graphs for support here.
% These are identical so we include only graphs for support here.
% We include only the graphs for support, because those for confidence are identical.

\begin{figure}[t]
\centerline{\includegraphics[width=\linewidth]{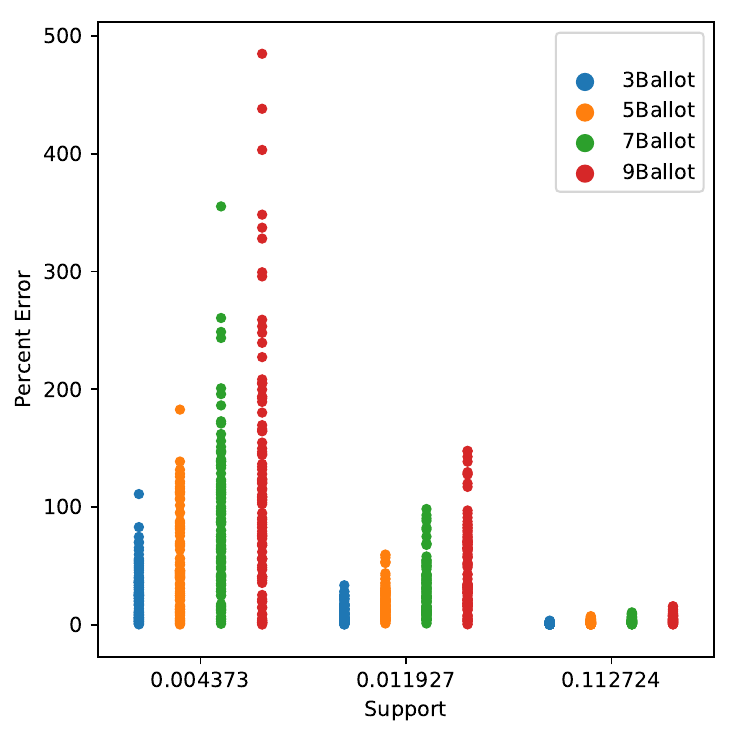}}
  \caption{Percent error for the support over two elements as rule occurrences vary in the case 3, 5, 7 and 9Ballot.}
  \label{fig:percent_error_pos}
\end{figure}

Our first experiment studies the percent error for the support over two elements when varying the number of rule occurrences for a dataset of $1M$ records.
Figure~\ref{fig:percent_error_pos} shows the results in the case of 3, 5, 7 and 9Ballot.
Element sets that occur less often are prone to higher percent error, with a high variance in the reported support values. 
This is expected for rules with very low support as, for example, observing a rule twice in $D_{priv}$ when it occurs only once in $D$ gives a percent error of $100\%$ despite the practically meaningless difference.
As element sets become more frequent (up to around $11\%$), the percent error ($<2\%$) and the variance both shrink.
As the difference in percent error between MultiBallot schemes also shrinks we focus on the results for 3Ballot in the following experiments.

% \alex{Could include more in the appendix.}

\begin{figure}[t]
\centerline{\includegraphics[width=\linewidth]{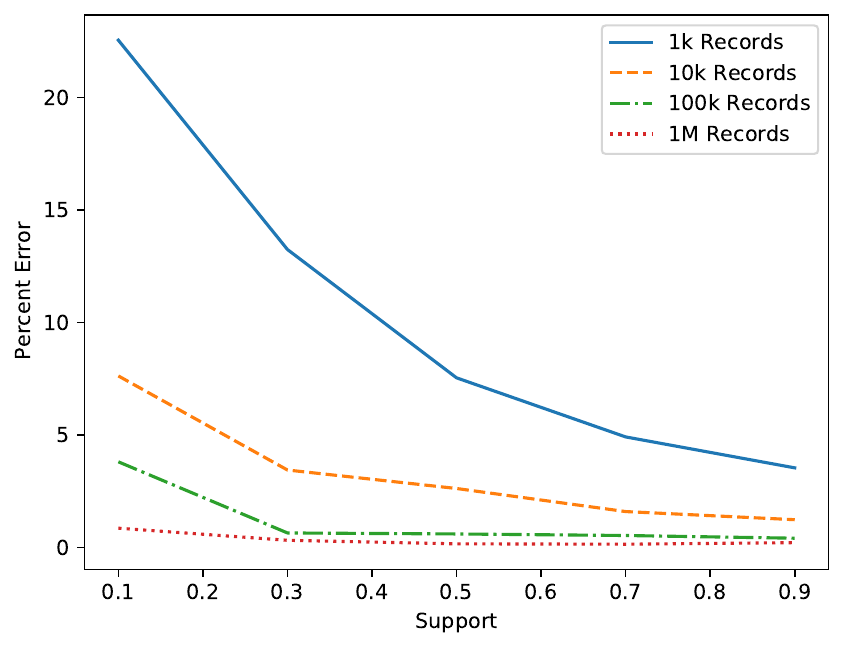}}
  \caption{Percent error for elements that appear with varying frequency in datasets with different number of users, using 3Ballot.}
  \label{fig:percent_error_users}
\end{figure}

Our second experiment studies whether the accuracy for an element set depends on the number of times the element set occurs, or its occurrences relative to the overall number of users i.e., support.
We generate four datasets of size $1k$, $10k$, $100k$, and $1M$, and pick five element sets with support $0.1$, $0.3$, $0.5$, $0.7$, and $0.9$, from each dataset.
The percent error (shown in Figure~\ref{fig:percent_error_users}) shrinks as the support increases, but the absolute size of the element set plays a bigger role in the accuracy of the statistics.
In the cases of the $100k$ and $1M$ user datasets, the support has only a minimal effect on the accuracy.
Our results are consistent with those of Blum et al~\cite{blum2005practical}.

\begin{figure}[t]
\centerline{\includegraphics[width=\linewidth]{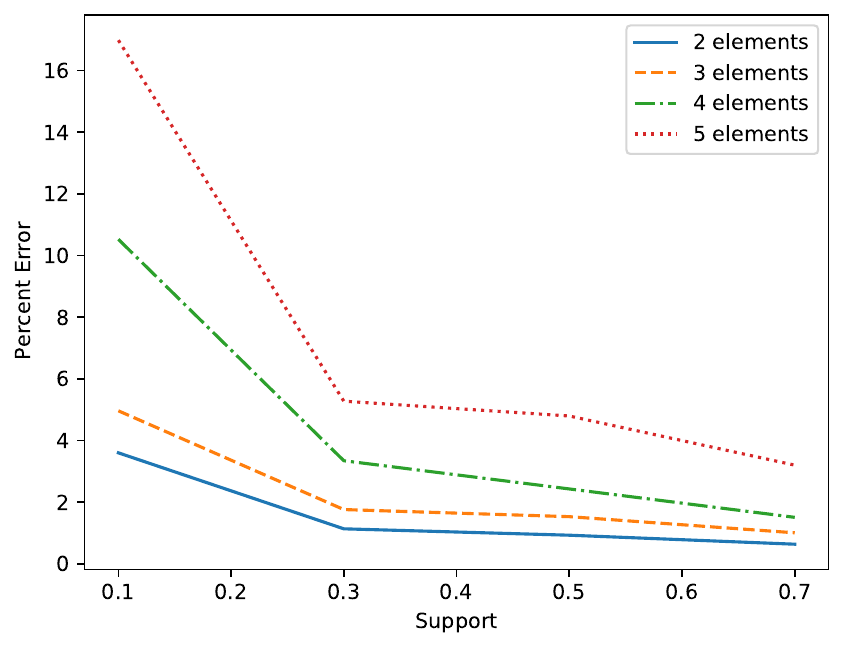}}
  \caption{Percent error for element sets of varying size that have the same support, using 3Ballot.}
  \label{fig:percent_error_traits}
\end{figure}

Our third experiment evaluates MultiBallot for different element set sizes using a synthetic dataset of $100k$ users.
Figure~\ref{fig:percent_error_traits} shows that accuracy is sensitive to increases in the number of elements.
This is expected as the scheme probabilistically estimates the field values of the original record based on the observed shares, and the inference error for each field adds up with the number of elements.

Our results show that MultiBallot can provide publicly verifiable statistics in the context of law-enforcement access to telecommunications data i.e., the type of statistics published by the IPCO~\cite{ipco2018}.

To evaluate the applicability to healthcare data, we consider two types of studies: studies on genes and protein networks, and epidemiology studies.
In studies on genes and protein networks, datasets commonly contain between $100k$ and a few million records, with a support threshold usually around $0.5\%$.
In most cases, valid association rules are composed of only two elements and their support is greater than the minimum threshold. (The minimum threshold is relevant only during the rule-mining phase.)
% This is important because the minimum threshold is relevant only during the rule mining phase.
In the verification phase, the users compute measures over the relationships that are reported by the researcher as strongly associated~\cite{guzzi2014mining,nagar2015association,faria2012mining,kumar2004dependence}.

In epidemiology studies, the average element set size is $3$, with a minimum support of around $1\%$.
However, the support of relevant element sets identified is much higher and ranges from $1\%$ to $16\%$, and datasets contain between $10,000$ and $250,000$ records~\cite{park2014association,jensen2012mining,toti2016analysis}.
Our analysis of MultiBallot shows that acceptable accuracy can be obtained for such studies of healthcare data.

% \subsection{Deployability}
% For a system like \sysname to be deployed, agents and data providers would need to implement the necessary infrastructure.
% They may do so to increase public confidence that their access to personal data is legitimate~\cite{deepmind}.
% % Participants may also implement \sysname to allow them to demonstrate that data they submit as evidence in legal proceedings has not been tampered.
% % 
% Alternatively, they may have a statutory obligation to provide transparency e.g., compliance with ETSI requirements may be a condition of providing a telecommunication service.
% Such standards do include provision for requiring that access to personal data is auditable and that the authenticity of data can be established~\cite{ETSI}.

% In the UK, the IPCO can require that public authorities and telecommunication operators provide the commissioner's office with any assistance required to carry out audits, and this could include implementing IT infrastructure~\cite[Section 235(2)]{IPAct}.
% Another possible route for imposing a statutory requirement to provide transparency could be through enforcement action of a regulator such as the Federal Trade Commission or a data protection authority.

%%%%%%%%%
\section{Deployability}
% \alex{Note here or in evaluation section that our benchmarks are performed on cheap af machines so deploying would be affordable}

For a system like \sysname to be deployed, agents and data providers would need to implement the necessary infrastructure.
They may do so as part of transparency initiatives to increase public confidence~\cite{deepmind}.
As we have shown in the benchmarks presented in Section~\ref{sec:implementation}, this may be cost-effective as VAMS can achieve good enough performance on very cheap hardware.

% They may do so to increase public confidence~\cite{deepmind} or due to an obligation to provide transparency e.g., compliance with ETSI requirements may be a condition of providing a telecommunication service.
Parties may also implement \sysname to allow them to demonstrate that data they submit as evidence in legal proceedings has not been tampered with.
Alternatively, they may have a statutory obligation to provide transparency e.g., compliance with ETSI requirements may be a condition of providing a telecommunication service.
Such standards do include provisions for requiring that access to personal data is auditable and that the authenticity of data can be established~\cite{ETSI}.

In the UK, the IPCO can require that public authorities and telecommunication operators provide the commissioner's office with any assistance required to carry out audits, and this could include implementing IT infrastructure~\cite[Section 235(2)]{IPAct}.
Another possible route for imposing a statutory requirement to provide transparency could be through enforcement action of a regulator such as the Federal Trade Commission or a data protection authority.
NGOs that currently work with transparency as an objective (e.g., make Freedom of Information requests) could also have an interest in maintaining and operating a system like \sysname by, for example, hosting log servers and serving as data brokers or auditors.

% \paragraph*{Implementing transparency}
% An important aspect of transparency is not only that information should be released, but also the utility of the information.
% In particular, the information should not be useful only for honest parties to show that they are honest.
% For example, if the information released cannot be verified (e.g., false statistics) and does not include individual outcomes to determine whether the system is fair, transparency will usually be ineffective or even counter-productive~\cite{taylor2016transparency}.

% Here, we have designed \sysname to provide effective transparency by enabling verifiable population and individual outcomes, which requires the release of data rather than reliance on cryptographic proofs, while maintaining a reasonable trade-off between transparency and privacy.

%%%%%%%%%
\section{Related Work}\label{sec:rel}
For a systematisation of transparency enhancing technologies based on cryptographic logs such as VAMS, see the work of Hicks~\cite{hicks2023sok}.

The work closest to ours is due to Frankle et al.~\cite{frankle2018practical}, who propose a system that allows accountability of secret legal processes using zero-knowledge proofs and aggregate statistics computed through a multi-party computation (MPC) between courts.
Previously, Goldwasser and Park~\cite{goldwasser2017public} had also proposed using append-only ledgers and zero-knowledge proofs in the context of actions related to secret laws under the U.S. Foreign Intelligence Surveillance Act (FISA).

This approach provides less transparency than ours as they do not support individual transparency, only aggregate statistics, thereby reducing the potential for users to contest outcomes~\cite{hicks2022transparency}.
While the outputs of zero-knowledge proofs and MPC can be checked for correctness, the integrity of inputs (i.e. the integrity of the data used in audits) cannot be verified.
Because the inputs can be manipulated, they must assume that judges (who are closest to the auditors of our context) are not malicious and would not publish an inaccurate report, making their threat model weaker than ours.
Their proposed systems are also specific to a targeted use case where all parties could coordinate to perform the required multiparty computation, while our approach is more generally applicable as parties do not require as much coordination aside from the initial request for data (which is unavoidable).

Work by Panwar et al.~\cite{panwar2019sampl} also addresses the problem of auditing surveillance orders, but differs from ours to a greater extent as it envisions an \textit{enforcer} that verifies the interactions between agents and data providers, which are recorded on a blockchain using zero-knowledge proofs, but does not support verifiable statistics.

% Crosby and Wallach~\cite{crosby2009efficient} propose an efficient construction of a log in the case of an untrusted logger serving clients storing events in a log kept honest through auditing.
% They use a centralized hash-tree based log (which inspired the one used in Certificate Transparency~\cite{CT}) but do not address secrecy of logged events or replication.
% We are also primarily concerned with the use of the log rather than the design of a log.
Tamper-evident logs have also been used in other work that focuses on auditability.
Bates et al.~\cite{bates2015accountable} look at accountable logs of wiretapping in the context of equipment implementing requirements of the US Communications Assistance for Law Enforcement Act (CALEA). This system permits simple counting queries, whereas \sysname allows broader analysis.
CONIKS~\cite{melara2015coniks} deals with the specific case of key transparency, allowing users to monitor their key bindings, and does not deal with other problems that we address, in particular public audits.

In terms of privacy preserving statistics, techniques such as k-anonymity~\cite{sweeney2002k}, l-diversity~\cite{li2007t}, t-closeness~\cite{machanavajjhala2006diversity} and $\rho$-uncertainty~\cite{cao2010rho} have been proposed.
As discussed by Domingo-Ferrer and Torra~\cite{domingo2008critique}, however, these techniques provide privacy only when the utility of the dataset is significantly reduced, whereas our solution enables accurate statistics.
% Thus, in most cases the anonymized dataset does not contain enough information to extract statistically significant insights about the population.

Another line of work that attempts to address this limitation is privacy-preserving association rule mining~\cite{agrawal1993mining} (we introduce association rule mining in Section~\ref{sec:system}).
Such techniques generate randomized or perturbed datasets that protect the privacy of users while preserving some of the associations between the variables that are of interest.
Originally, privacy-preserving association rule mining was performed through uniform randomization of the dataset based on a public factor.
As shown by Evfimievski et al.~\cite{evfimievski2004privacy} this naive approach does not protect the users' privacy effectively.
They instead proposed \textit{randomization operators}\cite{evfimievski2004privacy} that were also proven ineffective and require an initial dataset of at least one million records~\cite{zhang2004new}.
Zhang et al. proposed a scheme that considers the existing association rules when perturbing the data and, as a result, provides better privacy bounds~\cite{zhang2004new}.
Unfortunately, this scheme has limited applicability as it severely distorts the strength of the association rules, overestimating strong relationships and under-representing less frequent ones.
Overall, the weak privacy guarantees and the poor accuracy achieved by those schemes make them unsuitable for a system like \sysname.

A more promising line of work is based on differential privacy~\cite{diffpriv}.
Such schemes have been studied extensively in the past years and have been proven to be secure in a variety of settings~\cite{dpsurvey}.
However, they still impose trade-offs between privacy and utility~\cite{alvim2011differential}, as well as one-shot and continuous observation~\cite{dwork2010differential}.
Achieving a meaningful privacy parameter can also be hard in practice~\cite{lee2011much}, particularly when the aim is to provide a general solution like ours.
This problem is tackled by Chen et al.~\cite{chen2011publishing}, who take into consideration the underlying dataset to provide stronger privacy guarantees and increased utility.

None of these solutions provides verifiability, however, so the public cannot easily verify the integrity of the published data or statistics.
In fact, the analyst who adjusts the noise term may accidentally or intentionally sample from distributions that drastically skew the statistics computed~\cite{verdp}.

Narayan et al.~\cite{verdp} solve this problem with a scheme that uses a subset of Fuzz~\cite{HaeberlenPN11} to generate publicly verifiable validity proofs.
Unfortunately, VerDP has limited expressiveness and severely constrains access to the dataset.
More specifically, once the privacy budget of a particular dataset gets depleted, no further queries or analyses can be conducted.
This may exclude researchers from using the data and prevent the application of novel analysis techniques on older, depleted datasets.
It could also allow a malicious party to intentionally deplete the privacy budget.
In comparison, we allow the data to be used any number of times and without constraints.

\section{Conclusion}\label{sec:conclusions}
We have proposed a system, \sysname, which achieves our transparency and privacy goals.
Our work shows how existing transparency overlays used to provide tamper-evident logging can be combined with our log entry tagging scheme and MultiBallot to support publicly verifiable individual and population level transparency about access to data requests.
% generate differentially private synthetic data that allows statistics to be released and publicly verified without compromising privacy.
Our evaluation of two implementations of \sysname shows that the system also meets realistic performance requirements in practice, and not only on paper.

% An important aspect of transparency is not only that information should be released, but also the utility of the information.
% In particular, the information should not be useful only for honest parties to show that they are honest.
% For example, if the information released cannot be verified (e.g., false statistics) and does not include individual outcomes to determine whether the system is fair, transparency will usually be ineffective or even counter-productive~\cite{taylor2016transparency}.
% \sysname provides effective transparency by enabling verifiable population and individual outcomes.

Our results illustrate that the current framework for requesting data can be greatly improved to benefit all parties involved.
We have given two example use cases in Section~\ref{sec:introduction} to illustrate how \sysname could be used.
Its design does not depend on any particularities of these use cases so it could therefore be applied more generally.
\sysname does not have to replace any existing component in the workflow of an organization.
Instead, it serves as an overlay that can be used to achieve both transparency and privacy goals.

%%%%%%%%%%%%%%%%%%%%%%%%%%%%%%%%%%%%%%%%%%%%%%%%%%%%%%%%%%%%%%%%%%%%%%%%%%%%%%%%%%%%%%%%%%%%%%%%%%%%%%%%%%%%
%%%%%%%%%%%%%%%%%%%%%%%%%%%%%%%%%%%%%%%%%%%%%%%%%%%%%%%%%%%%%%%%%%%%%%%%%%%%%%%%%%%%%%%%%%%%%%%%%%%%%%%%%%%%
%%%%%%%%%%%%%%%%%%%%%%%%%%%%%%%%%%%%%%%%%%%%%%%%%%%%%%%%%%%%%%%%%%%%%%%%%%%%%%%%%%%%%%%%%%%%%%%%%%%%%%%%%%%%
\section*{Acknowledgments}
The authors would like to thank Jonathan Bootle, Paul Dunphy, and Wai Yi Feng for helpful discussions and suggestions.
Alexander Hicks was supported by OneSpan\footnote{\url{https://www.onespan.com/}} and  UCL  through  an  EPSRC  Research  Studentship, Vasilis Mavroudis was supported by the European Commission through the H2020-DS-2014-653497 PANORAMIX project, Mustafa Al-Bassam was supported by a scholarship from The Alan Turing Institute, Sarah Meiklejohn was supported in part by EPSRC Grant EP/N028104/1 and in part by a Google Faculty Award, and Steven Murdoch was supported by The Royal Society [grant number UF160505].
% %-------------------------------------------------------------------------------
% \section*{Availability}
% %-------------------------------------------------------------------------------

% Our code will be made available upon publication to avoid deanonymisation.
% %-------------------------------------------------------------------------------

\bibliographystyle{plain}
\bibliography{refs}

\end{document}